\documentclass[aps,prd,twocolumn,superscriptaddress,nofootinbib]{revtex4-2}

\usepackage{amsmath}
\usepackage{amssymb}
\usepackage{graphicx}
\usepackage{float}
\usepackage{multirow}
\usepackage{hyperref}
\usepackage{xcolor}
\usepackage{array}
\usepackage{dcolumn}
\usepackage{mathtools}
\usepackage{orcidlink}
\usepackage[normalem]{ulem}

\newcommand{\e}{\varepsilon}
\newcommand{\beq}{\begin{equation}}
\newcommand{\eeq}{\end{equation}}
\newcommand{\gexact}{{\sf g}}

\begin{document}

\title{Self-Force Calculations with a Spinning Secondary}

\author{Josh Mathews\,\orcidlink{0000-0002-5477-8470}}
\affiliation{School of Mathematics \& Statistics, University College Dublin, Belfield, Dublin 4, Ireland, D04 V1W8}

\author{Adam Pound\,\orcidlink{0000-0001-9446-0638}}
\affiliation{School of Mathematical Sciences and STAG Research Centre, University of Southampton, Southampton, United Kingdom, SO17 1BJ}

\author{Barry Wardell\,\orcidlink{0000-0001-6176-9006}}
\affiliation{School of Mathematics \& Statistics, University College Dublin, Belfield, Dublin 4, Ireland, D04 V1W8}

\begin{abstract}

We compute the linear metric perturbation to a Schwarzschild black hole generated by a spinning compact object, specialising to circular equatorial orbits with an (anti-)aligned spin vector. We derive a two-timescale expansion of the field equations, with an attendant waveform-generation framework, that includes all effects through first post-adiabatic order, and we use the Regge-Wheeler-Zerilli formalism in the frequency domain to generate waveforms that include the complete effect of the spin on the waveform phase. We perform the calculations using expansions at fixed orbital frequency, increasing the computational efficiency and simplifying the procedure compared to previous approaches. Finally, we provide the first fully relativistic, first-principles regularisation procedure for gauge invariant self-force quantities to linear order in spin. We use this procedure to produce the first \textit{strong-field, conservative} self-force calculation including the spin of the secondary --- computing Detweiler's redshift invariant.

\end{abstract}

\maketitle

\section{Introduction}
\label{sec:introduction}

The era of gravitational wave astronomy was beckoned by the first detection of gravitational waves by the LIGO and VIRGO collaborations \cite{LIGOScientific:2016aoc}. The collaborations' ground-based detectors now frequently detect signals from the mergers of binary systems of neutron stars and stellar-mass black holes \cite{LIGOScientific:2018mvr}. The advent of future space-based gravitational wave detectors will bring the ability to observe gravitational waves in lower frequency bands. The European Space Agency is currently scheduled to launch LISA, a space-based gravitational wave detector, in 2034 \cite{LISA:2017pwj}. Extreme mass ratio inspirals (EMRIs) are an important source of gravitational waves for LISA, with frequencies optimally placed in LISA’s sensitivity band \cite{Babak:2017tow}. EMRIs consist of a compact `secondary body' (of mass $\mu$) such as a stellar mass black hole or neutron star, orbiting a significantly more massive `primary body' (of mass $M$) such as a supermassive black hole. The system radiates energy as gravitational waves, leading the secondary's orbit to gradually inspiral before it eventually collides and merges with the primary body. EMRI configurations evolve slowly while the secondary body effectively surveys the primary's spacetime. 
This information is imprinted on the resulting gravitational waveforms, including data in the strong gravitational field as the separation of the bodies diminishes. Thus the scientific potential of EMRI detection is particularly alluring and can be used as an accurate test of General Relativity in the strong field, as well as giving insight into compact object populations \cite{Babak:2017tow}. The detection of EMRI waveforms and parameter estimation of the binary's physical attributes will require precise waveform models.

The extreme mass ratio of EMRI systems can be exploited by modelling their waveforms using the self-force approach of black hole perturbation theory (see for example, \cite{pound2021black, Barack:2018yvs,  PPVLivingReview}). In this approach, the primary body's spacetime is perturbed by the secondary and the resulting perturbed spacetime metric is expanded in powers of the mass ratio, $\mu/M$. The Einstein field equations can then be solved order by order in the mass ratio for the metric perturbations, which in turn can be used to calculate EMRI inspirals and waveforms. Calculations involving the first order in mass ratio metric perturbations are referred to as first-order self-force (1SF) calculations, while calculations involving the second-order metric perturbations are referred to as second-order self-force (2SF) calculations, and so on.

Leading order approximate EMRI waveforms, known as `adiabatic' waveforms, require only partial 1SF calculations. The phase evolution of these adiabatic waveforms is obtained from the 1SF \textit{dissipative} gravitational self-force. The remaining \textit{conservative} 1SF contributes to the waveform at next-to-leading order or \textit{first post-adiabatic} order (1PA). The other contributions to 1PA waveforms are the dissipative part of 2SF (without spin effects) and terms related to the spin of the secondary body. These additional spin terms comprise of the spin-curvature coupling contributions and the linear in spin, linear in mass ratio dissipative self-force and self-torque. Detection of weaker EMRI signals and accurate parameter estimation necessitates 1PA waveform modelling. Practical calculations of 1PA waveforms take advantage of the slow evolution of EMRI configurations compared to the timescale of the orbital motion of the secondary --- the two-timescale expansion~\cite{Miller:2020bft,pound2021black}. In this paper we develop a two-timescale expansion including all spin effects through 1PA order, in the special case of quasicircular orbits and (anti-)aligned spin. This expansion provides a practical method of solving the field equations, a rapid, first-principles waveform-generation framework, and a simple, modular way of including spin effects into 1PA waveform calculations.  When combined with recent progress on 2SF calculations  \cite{Pound:2019lzj, Warburton:2021kwk}, this will enable production of complete 1PA waveforms, including (anti-)aligned spin for quasicircular inspirals into non-spinning (or slowly spinning~\cite{Warburton:2021kwk}) black holes.

Although the conservative linear in spin self-force only contributes to the waveform at post-2-adiabatic order (and is thus not required for 1PA waveform modelling), the complete leading order self-force including the leading spin effects is interesting in its own right. Conservative self-force dynamics can be used to inform the effective-one-body formalism \cite{Buonanno:1998gg} or `EOB' for short (see \cite{Damour:2009sm, Barack:2010ny, Barausse:2011dq, Akcay:2012ea, Antonelli:2019fmq, Bini:2019nra, Khalil:2021fpm}), especially with the discovery of the `First Law of Binary Black Hole Mechanics' \cite{LeTiec:2011ab} and the development of powerful scattering techniques. As detector sensitivity improves, it will become increasingly critical to make use of this SF information to improve EOB models of spin effects, eccentricity, and small mass ratios.
Useful ingredients to do so are gauge invariant quantities such as Detweiler's redshift invariant --- first introduced for non-spinning secondary bodies in circular equatorial orbits in Schwarzschild spacetime in \cite{Detweiler2008} --- and the spin precession invariant \cite{Dolan:2013roa}. The definition of the redshift invariant was generalised to include spin effects in \cite{Bini:2018zde}, and in the same work a Post-Newtonian (PN) approximation was calculated for the case of an aligned-spin secondary in a circular equatorial orbit around a Schwarzschild primary. This calculation was later extended to include the body's spin-induced quadrupole moment \cite{PhysRevD.102.024091}. There have been several calculations of the spin precession invariant \cite{Dolan:2013roa, Akcay:2016dku, Kavanagh:2017wot, Bini:2019lkm}, though to our knowledge these are yet to be extended to include the spin of the secondary.

In black hole perturbation theory, a small extended `test' body experiences an acceleration driven by perturbation forces induced by the body's own multipole moments coupling to the background spacetime. The pole-dipole approximation for a spinning test body assumes the body can be described by its mass-monopole and spin-dipole moments only, neglecting the quadrupole and higher moments\footnote{Note that the term `multipoles' is also used separately to refer to spherical harmonic modes (such as `the low multipoles' in completing the metric perturbation).}. Omitting perturbations to the background spacetime, the motion of a spinning test body is governed by the Mathisson-Papetrou-Dixon (MPD) equations \cite{Mathisson1937,Papapetrou1951,Dixon1970} with a supplementary spin condition (SSC). The MPD equations assume that the length scale associated with the spin of the test body are much smaller than the length scale of the curvature of the background spacetime --- this leads us to include only linear in spin effects. A thorough study of motion governed by the MPD equations was produced in \cite{Witzany:2018ahb, Witzany:2019nml}. The consistency of different physical results for solutions of the MPD equations with several common SSCs was investigated in \cite{Harms:2016ctx, PhysRevD.104.024042}. The results in \cite{PhysRevD.104.024042} showed, for example, that varying the SSC can change the expression for the orbital frequency of a spinning body in a circular orbit at quadratic and higher order in spin. As the secondary's spin is proportional to the square of its mass, the leading linear spin terms enter the metric perturbation at second order in the mass ratio, and linearizing in spin is necessary to consistently satisfy the equations of motion at each order in the mass ratio. Defining the dimensionless spin parameter $\chi\equiv S/\mu^2$ where $S$ is the magnitude of the spin vector, $S^2\equiv S^{\beta}S_{\beta}=\frac{1}{2} S_{\mu \nu} S^{\mu \nu}$, this motivates us to work with the dimensionless spin tensor, $\tilde S^{\mu\nu} \equiv S^{\mu\nu}/\mu^2$. If the compact secondary is a black hole, for example, then $|\chi| \leq 1$. The spin-linearized metric perturbation has the additional advantage that it can be computed without fixing a value of $\chi$, making it a very efficient way of filling the parameter space of $\chi$ values.

First order gravitational self-force calculations began by examining specific EMRI models with a non-spinning secondary body and progressed towards generic EMRI configurations \cite{vandeMeent:2017bcc}. Recently, there has been progress on incorporating the secondary's spin in specific EMRI configurations. The effect of the spin-curvature coupling (MPD force) was incorporated into self-force calculations in \cite{PhysRevD.96.084057} for a Schwarzschild black hole primary, and waveforms were produced including the coupling for secondaries in eccentric equatorial motion with the spin vector aligned to the orbital angular momentum. That work neglected the spin contribution to the gravitational wave fluxes. The spin-curvature coupling in a Kerr background spacetime was studied in detail in the frequency domain in \cite{Ruangsri:2015cvg} with a view to informing perturbative EMRI calculations. Fluxes for a Kerr black hole primary and an aligned spinning secondary in a circular equatorial orbit were produced in the frequency domain in \cite{Han:2010tp} and in the time domain in \cite{Harms:2015ixa}, without truncating quadratic and higher order in spin terms. A flux balance law was derived in \cite{Akcay:2019bvk} that holds for generic orbits in Kerr to linear order in spin and the law was demonstrated for aligned-spin circular equatorial orbits in Schwarzschild spacetime, truncating at linear order in spin. Linearized-in-spin fluxes were computed for a secondary in a spin-aligned circular equatorial orbit around a Kerr primary in \cite{Piovano:2020zin} and used to calculate waveforms. That work was later extended to investigate aspects of the detectability of the spin of the secondary in \cite{Piovano:2021iwv}. Flux calculations including spin effects for the most general EMRI configuration that we are aware of were performed in \cite{PhysRevD.103.104045}, with a Kerr primary and an aligned-spinning secondary in an eccentric equatorial orbit, without truncating quadratic in spin and higher terms.

Of all the self-force flux calculations including spin effects, we are not aware of any that generalise the spin of the secondary away from the (anti-)aligned case. The numerical calculations in this work will continue this legacy, specialising to secondaries with (anti-)aligned spin in circular equatorial orbits around a Schwarzschild primary black hole. We calculate both the dissipative and conservative leading self-force including the next-to-leading-order spin effects, and derive from first principles a fully relativistic regularization procedure for gauge invariant self-force quantities to linear order in spin. In doing so, we provide a covariant expansion of the corresponding singular field. We also introduce a fixed frequency parameterisation where the (quasi)circular motion of the spinning body can be parameterised by a spin \textit{independent} frequency. In frequency domain calculations, this has a distinct advantage in linearizing numerical results in spin --- the numerically integrated perturbation equations depend on the frequency and if there is spin dependence in the frequency, then one must find a way to truncate numerical results at linear order in spin. Typically this has been done by numerically fitting the spin dependence \cite{Akcay:2019bvk, Piovano:2020zin} (significantly increasing the computational cost) or more recently by linearizing the perturbation equations and solving a coupled system with an extended source \cite{Piovano:2021iwv}. Removing the spin dependence from the frequency (and therefore from the numerical integration) entirely avoids the issues that come with a non-compact source, and produces results that are readily comparable with their post-Newtonian equivalents (calculated at fixed frequency) and that are simple to implement in the two-timescale expansion for producing waveforms.

In this paper we adopt the metric signature $-+++$ and work with geometrized units such that $G=c=1$. Symmetrisation of indices is denoted using parentheses while square brackets represent anti-symmetrisation. We use $\e$ as a formal order-counting parameter to keep track of powers of the mass ratio. Finally, when referring to gauge invariance, we mean invariance within a class of gauges that do not interfere with the helical symmetry of the perturbed spacetime.

In Section~\ref{sec:eom} we review the equations of motion for a spinning secondary body. In Section~\ref{sec:circular equatorial orbits} we specialise the motion to circular equatorial orbits in Schwarzschild spacetime with the secondary body's spin vector (anti-)aligned to its total angular momentum. In Section~\ref{sec:tt expansion} we develop the two-timescale expansion of the field equations and equations of motion. In Section~\ref{sec:rw} we review the Regge-Wheeler-Zerilli formalism for solving the perturbation equations to obtain the leading order metric perturbation (including sub-leading spin effects) and the gravitational energy and angular momentum fluxes associated with a spinning body. In Section~\ref{sec:redshift} we review Detweiler's redshift invariant for a spinning secondary. In Section~\ref{sec:reg} we derive a regularisation procedure for the treatment of the singular field associated with a spinning point particle and produce regularisation parameters for the redshift invariant. In Section~\ref{sec:results} we present our numerical results and some gravitational waveforms.

\section{Equations of motion with a spinning secondary}
\label{sec:eom}

\subsection{Mathisson-Papapetrou-Dixon-Harte equations}
\label{sec:mpd}

The equations of motion for a generic, spinning compact body are well established through order $\e$~\cite{PPVLivingReview} but only partially known at order $\e^2$. The most far-reaching results are due to Harte~\cite{Harte:2011ku}, who showed that in fully nonlinear gravity, a self-gravitating material body obeys the same equations of motion as a test body, but a test body in an {\em effective} metric $\hat g_{\alpha\beta}$. If we ignore the effects of the body's quadrupole and higher moments, then these test-body equations, referred to as the Mathisson-Papapetrou-Dixon (MPD) equations~\cite{Mathisson1937,Papapetrou1951,Dixon1970}, take a simple form:
\begin{subequations}\label{MPD eqns}
\begin{align}
\frac{\hat D p^{\alpha}}{d \hat\tau} &=-\frac{1}{2} \hat R^{\alpha} {}_{\beta \gamma \delta}\hat u^{\beta} S^{\gamma \delta} ,\\
\frac{\hat D S^{\gamma \delta}}{d \hat\tau} &=2 p^{[\gamma} \hat u^{\delta]},
\end{align}
\end{subequations}
where $p^{\mu}$ is the body's linear momentum, $\hat u^\mu=dz^\mu/d\hat\tau$ is its four-velocity, $S^{\gamma \delta}$ is its spin tensor, and $\hat D/d\hat \tau = \hat u^\alpha\hat\nabla_\alpha$. $\hat \tau$ and $\hat \nabla$ are the proper time and covariant derivative compatible with $\hat g_{\alpha\beta}$. Several decades earlier, Thorne and Hartle~\cite{Thorne-Hartle:85} showed that these equations also hold for a black hole. However, the results of both Harte and Thorne and Hartle (particularly the latter) are limited by an incomplete characterization of $\hat g_{\alpha\beta}$. One of our ancillary goals will be to partly solidify their status.

For any given $\hat g_{\alpha\beta}$, Eq.~\eqref{MPD eqns} is an underdetermined system of 10 equations for 13 unknowns: the six components of  the antisymmetric tensor $S^{\gamma \delta}$, the 4 components of $p^{\mu}$, and the three independent components of $\hat u^{\nu}$. A ``spin supplementary condition'' is therefore required to uniquely determine the solution. Common spin supplementary conditions are $\hat g_{\alpha\beta}\hat u^\alpha S^{\beta\gamma}=0$ and $\hat g_{\alpha\beta}p^\alpha S^{\beta\gamma}=0$. If we work to linear order in the spin, these conditions are equivalent, and they imply $p^\alpha=\mu \hat u^\alpha$. The MPD equations in $\hat g_{\alpha\beta}$ then reduce to
\begin{subequations}\label{SSC MPD eqns}
\begin{align}
\label{eqn:MPD}
\frac{\hat D^{2} z^{\mu}}{d \hat\tau^{2} }= \frac{\hat D \hat u^{\alpha}}{d \hat\tau} &=-\frac{\mu}{2 } \hat R^{\alpha}{}_{\beta \gamma \delta} \hat u^{\beta} \tilde S^{\gamma \delta}, \\
\label{eqn:MPD2}
\frac{\hat D \tilde S^{\gamma \delta}}{d \hat\tau} &= 0,
\end{align}
\end{subequations}
where we have used $S^{\mu\nu}=\mu^2\tilde S^{\mu\nu}$.
  
\subsection{Effective metric, self-force, and self-torque}
\label{sec:gsf}

In self-force theory, the physical metric of the spacetime is expanded in the form 
\begin{equation}\label{generic expansion of g}
  \mathbf{g}_{\alpha \beta}  = g_{\alpha \beta} + \e h^{(1)}_{\alpha \beta} + \e^2 h^{(2)}_{\alpha \beta} + \mathcal{O}(\e^3),
\end{equation}
where $\e$ is an order-counting parameter that counts powers of $\mu$. In our context, $g_{\alpha \beta}$ is the metric of the central black hole and $h_{\alpha\beta}\equiv\sum \e^n h^{(n)}_{\alpha\beta}$ is the perturbation due to the presence of the secondary. Since $S^{\alpha\beta}\sim\mu^2$, the secondary's spin contributes (linearly) to $h^{(2)}_{\alpha \beta}$. Higher moments would contribute at order $\e^3$ and above.  

The effective metric is given by $\hat g_{\alpha\beta} = g_{\alpha\beta}+h^{\rm R}_{\alpha \beta}$, where $h^{\rm R}_{\alpha \beta}=\sum\e^n h^{{\rm R}(n)}_{\alpha\beta}$ is a certain piece of the physical perturbation $h_{\alpha\beta}$. We can expand the MPD equations~\eqref{SSC MPD eqns} in powers of $h^{\rm R}_{\alpha \beta}$ following Sec.~IIIA of Ref.~\cite{Pound:2015fma} (for example). Defining the difference between the connections on $\hat g_{\alpha\beta}$ and $g_{\alpha\beta}$ as $C^\alpha{}_{\beta\gamma} \equiv\hat \Gamma^\alpha_{\beta\gamma}-\Gamma^\alpha_{\beta\gamma}$, we have the standard relations
\begin{align}
C^\alpha{}_{\beta\gamma} &= \frac{1}{2}\hat g^{\alpha\delta}\left(2h^{\rm R}_{\delta(\beta;\gamma)}-h^{\rm R}_{\beta\gamma;\delta}\right),\\
\hat R^\alpha{}_{\beta\gamma\delta} &= R^\alpha{}_{\beta\gamma\delta} + 2C^\alpha{}_{\beta[\delta;\gamma]}+2C^\alpha{}_{\rho[\gamma}C^\rho{}_{\delta]\beta},
\end{align}
where a semicolon denotes the covariant derivative compatible with $g_{\alpha\beta}$. The proper times in the two metrics are related by $d\hat\tau/d\tau = \sqrt{1-h^{\rm R}_{\alpha\beta}u^\alpha u^\beta}$, where $u^\alpha = dz^\alpha/d\tau$. Substituting these relations into Eq.~\eqref{SSC MPD eqns} and expanding, we obtain\footnote{This corrects the analogous equations in Ref.~\cite{Akcay:2019bvk}. Since those equations were the starting point for Ref.~\cite{Akcay:2019bvk}'s derivation of the flux-balance law, we have independently re-derived the balance law, including the key intermediate result~\eqref{<dXidtau>}.}
\begin{subequations}\label{selfforceEOMs}
\begin{align}
\frac{D^{2} z^{\mu}}{d \tau^{2}} &= -\frac{1}{2} P^{\mu\nu}(g_\nu{}^\lambda - h^{\mathrm{R}\, \lambda}_\nu)\left(2 h_{\lambda \rho ; \sigma}^{\mathrm{R}}-h_{\rho \sigma ; \lambda}^{\mathrm{R}}\right) u^{\rho} u^{\sigma}\nonumber\\
&\quad -\frac{\mu}{2}R^{\mu}{}_{\alpha \beta \gamma}\left(1-\frac{1}{2}h^{\mathrm{R}}_{\rho\sigma}u^\rho u^\sigma\right)u^{\alpha}\tilde S^{\beta \gamma} \nonumber\\
&\quad +\frac{\mu}{2} P^{\mu\nu}(2h^{\mathrm{R}}_{\nu(\alpha;\beta)\gamma}-h^{\mathrm{R}}_{\alpha\beta;\nu\gamma})u^{\alpha}\tilde S^{\beta \gamma} +\mathcal{O}(\e^3)\nonumber\\
&\equiv F^\mu, \label{selfforceorbit}\\
\frac{D\tilde S^{\mu\nu}}{d \tau} &= u^{(\rho}\tilde S^{\sigma)[\mu}g^{\nu]\lambda}\left(2 h_{\lambda \rho ; \sigma}^{\mathrm{R}}-h_{\rho \sigma ; \lambda}^{\mathrm{R}}\right)  + \mathcal{O}(\e^2)\nonumber\\
&\equiv N^{\mu\nu},\label{selfforcespin}
\end{align}
\end{subequations}
where $P^{\mu\nu}\equiv g^{\mu\nu}+u^\mu u^\nu$. The spin-independent terms in $F^\mu$ are referred to as the self-force (per unit $\mu$) and $N^{\mu\nu}$ as the self-torque (per unit $\mu^2$). Note that Eq.~\eqref{selfforcespin} is expanded to one order lower than Eq.~\eqref{selfforceorbit} because the spin itself enters into the metric at one order higher than the trajectory $z^\mu$.

We can also extract an evolution equation for the scaled spin vector $\tilde{S}^{\mu} = -\frac{1}{2} \epsilon^{\mu}{}_{\alpha \beta \gamma} u^{\alpha} \tilde{S}^{\beta \gamma}$. Substituting $\tilde{S}^{\mu \nu}=-\epsilon^{\mu \nu}{}_{\alpha \beta}\tilde{S}^{\alpha}u^{\beta}$ into Eq.~\eqref{selfforcespin} and contracting the equation with $P^\alpha_{\ \beta}$ (to project out components tangent to $u^\alpha$), we find
\beq
P^\alpha_{\ \beta} \frac{D\tilde S^\beta}{d\tau} = N^\alpha, 
\eeq
where $N^\mu\equiv -\frac{1}{2}\epsilon^\mu{}_{\nu\rho\sigma}u^\nu N^{\rho\sigma}$.

The secondary's spin contributes to the above equations in three ways: through the standard MPD terms, which are independent of $h^{\rm R}_{\mu\nu}$; through terms of the form $h^{\rm R}\cdot S$, which can be considered as spin-induced self-forces and self-torques; and by contributing to $h^{{\rm R}(2)}_{\mu\nu}$ (via the spin's contribution to $h^{(2)}_{\mu\nu}$). We can write these contributions explicitly as
\begin{subequations}\label{spin F and N}
\begin{align}
F^\mu_{(\chi)} &= -\frac{\e^2}{2} P^{\mu\nu}\left(2 h_{\nu \rho ; \sigma}^{\mathrm{R}(\chi)}-h_{\rho \sigma ; \nu}^{\mathrm{R}(\chi)}\right) u^{\rho} u^{\sigma}\nonumber\\
&\quad -\frac{\mu}{2}R^{\mu}{}_{\alpha \beta \gamma}\left(1-\frac{\e}{2}h^{\mathrm{R}(1)}_{\rho\sigma}u^\rho u^\sigma\right)u^{\alpha}\tilde S^{\beta \gamma} \nonumber\\
&\quad +\frac{\mu\e}{2} P^{\mu\nu}(2h^{\mathrm{R}(1)}_{\nu(\alpha;\beta)\gamma}-h^{\mathrm{R}(1)}_{\alpha\beta;\nu\gamma})u^{\alpha}\tilde S^{\beta \gamma} +\mathcal{O}(\e^3), \label{spin force}\\
N^\mu_{(\chi)} &= \e u^{(\rho}\tilde S^{\sigma)[\mu}g^{\nu]\lambda}\left(2 h_{\lambda \rho ; \sigma}^{\mathrm{R}(1)}-h_{\rho \sigma ; \lambda}^{\mathrm{R}(1)}\right)  + \mathcal{O}(\e^2),\label{spin torque}
\end{align}
\end{subequations}
where we introduce $h^{{\rm R}(\chi)}_{\mu\nu}$ as the linear-in-spin piece of $h^{{\rm R}(2)}_{\mu\nu}$.

However, we note that the system of equations~\eqref{selfforceEOMs} is incomplete for two reasons: it omits terms of the same order as it keeps, specifically test-body spin-squared and quadrupole terms that first appear at $\mathcal{O}(\e^2)$; and we have not defined $h^{\rm R}_{\mu\nu}$. Since we restrict  our attention in this paper to linear spin effects, we freely skip over the first omission. The definition of $h^{\rm R}_{\mu\nu}$ is thornier. Harte defines a  class of effective metrics in which the MPD equations hold, but the specific choice he makes becomes singular on $z^\mu$ at order $\e^2$ in a perturbative expansion.\footnote{This can be deduced from Eqs.~(82)--(84) in Ref.~\cite{Harte:2011ku}, which express $\hat g_{\mu\nu}$ as the solution to $\hat g_{\mu\nu}=\mathbf{g}_{\mu\nu}-(\hat g_{\mu\rho}\hat g_{\nu\sigma}-\tfrac{1}{2}\hat g_{\mu\nu}\hat g_{\rho\sigma})H^{\rho\sigma}$, where $H^{\rho\sigma}$ is a solution to a linear differential equation in the spacetime $\hat g_{\mu\nu}$. For $\hat g_{\mu\nu}$ to be regular on the particle's worldline, $H^{\mu\nu}$ must cancel any singularities in $\mathbf{g}_{\mu\nu}$. This is impossible beyond linear order because if we substitute $\hat g_{\mu\nu} = g_{\mu\nu}+\e h^{\mathrm{R}(1)}_{\mu\nu}+\e^2 \hat g^{(2)}_{\mu\nu}+\mathcal{O}(\e^3)$ and Eq.~\eqref{generic expansion of g}, we find that $\hat g^{(2)}_{\mu\nu}$ cannot contain terms any more singular than $\sim h^{(1)}_{\mu\nu}h^{\mathrm{R}(1)}_{\rho\sigma}$, while $h^{(2)}_{\mu\nu}$ contains terms as singular as $h^{(1)}_{\mu\nu}h^{(1)}_{\rho\sigma}$.} Thorne and Hartle effectively define $\hat g_{\mu\nu}$ as the metric that exerts tidal fields on the body, leaving it open how to determine the body's own contribution to those fields. In Ref.~\cite{Pound:2012dk}, one of us (AP) defined a $\hat g_{\mu\nu}$ that is a smooth vacuum metric on $z^\mu$ and is well defined for both material bodies and black holes. For a nonspinning body, the equation of motion~\eqref{selfforceorbit} is valid for AP's definition of $\hat g_{\mu\nu}$~\cite{Pound:2012nt,Pound:2017psq}. But in the case of a spinning body, it has not been shown that the $O(\e^2)$ and $O(\e)$ terms in Eqs.~\eqref{selfforceorbit} and \eqref{selfforcespin} (respectively) are correct for AP's $\hat g_{\mu\nu}$. 

To skirt this issue, we note that Harte's definition is well defined and smooth for the effects we focus on: linear effects, whether linear in $\mu$ or linear in spin. For those contributions to $h^{{\rm R}(1)}_{\mu\nu}$ and $h^{{\rm R}(2)}_{\mu\nu}$, Harte's definition reduces to the more familiar Detweiler-Whiting definition~\cite{Detweiler2002}.\footnote{The only quantity in~\eqref{selfforceEOMs} that is not covered by this definition is the spin-independent piece of $h^{{\rm R}(2)}_{\mu\nu}$, for which one can use AP's definition.} We can therefore assume with some confidence that the linear-in-spin terms in~\eqref{selfforceEOMs} are correct with this definition of $h^{{\rm R}}_{\mu\nu}$.

We return to the regular field in Sec.~\ref{sec:reg}. There, as a byproduct of our concrete calculations, we show that Harte's definition of $h^{{\rm R}(\chi)}_{\mu\nu}$ agrees with AP's, giving us additional confidence in our assumption.

\subsection{Stress-energy of a spinning body}
\label{sec:stress-energy}

Since we are only interested in the gravitational field on scales much larger than the body's size, we can replace the body with a ``gravitational skeleton''~\cite{Mathisson1937}---a point singularity equipped with the body's multipole moments. Concretely, through second order in $\e$, a generic compact object can be modelled as a spinning point particle in $\hat g_{\mu\nu}$, with a stress-energy tensor 
\begin{equation}
\label{split}
T^{\alpha \beta}=\e T_{(\mu)}^{\alpha \beta} + \e^2 T_{(\chi)}^{\alpha \beta} + \mathcal{O}(\e^3),
\end{equation}
where $T_{(\mu)}^{\alpha \beta}$ is a mass-monopole term and $T_{(\chi)}^{\alpha \beta} $ is a spin-dipole term. Explicitly, the two contributions are
\begin{subequations}
\begin{align}
\label{stressenergy}
T_{(\mu)}^{\alpha \beta}(x) &=\mu\int d \hat\tau' \, \frac{\delta^{4}\left[x^{\mu}-z^{\mu}(\tau' )\right]}{\sqrt{-\hat g'}} \hat u^{\alpha}(\tau') \hat u^{\beta}(\tau' ),\\
\label{stressenergyspin}
T_{(\chi)}^{\alpha \beta}(x) &= \nabla_{\rho} \left[ \int d \tau'  \, \frac{\delta^{4}\left[x^{\mu}-z^{\mu}(\tau' )\right]}{\sqrt{-g'}} u^{(\alpha}(\tau' ) {S}^{\beta)\rho}(\tau') \right],
\end{align}
\end{subequations}
where $\delta^{4} $ is the four-dimensional Dirac delta function, and the covariant derivative $\nabla_{\rho}$ is with respect to the arbitrary field point $x^{\mu}$ and not the worldline point $z^{\mu}$.  Note that both $T_{(\mu)}^{\alpha \beta} $ and $T_{(\chi)}^{\alpha \beta}$ have subleading dependence on $\chi$ (and on $h^{\rm R}_{\alpha\beta}$) via their dependence on the worldline. However, the quantity $h^{\mathrm{R}(\chi)}_{\alpha\beta}$ defined in the previous section corresponds only to the regular field associated with $T_{(\chi)}^{\alpha \beta}$. Also note that no hats appear on quantities in $T_{(\chi)}^{\alpha \beta}$ since the difference would only contribute at order $\e^3$.

Unlike the equations of motion~\eqref{selfforceEOMs}, the stress-energy tensor~\eqref{split} has been rigorously derived from the method of matched asymptotic expansions~\cite{Pound:2012dk,Upton:2021oxf}. It holds for black holes, material bodies, and exotic compact objects.

\section{Circular orbits and aligned spins: test-spin effects}
\label{sec:circular equatorial orbits}

Before considering the full system of equations~\eqref{selfforceEOMs}, we consider the orbit and spin with $h_{\mu\nu}$ set to zero. The equations of motion are then given by the MPD equations Eq.~\eqref{SSC MPD eqns} with all hats removed; the orbit is accelerated in $g_{\mu\nu}$ by the MPD spin force on the right-hand side of Eq.~\eqref{eqn:MPD}, and the spin is parallel-propagated in $g_{\mu\nu}$. The results for this case will carry over directly to the full problem.

\subsection{Fixed-frequency parametrisation}

We specialise to the case of a non-spinning black hole primary, in which case the background is the Schwarzschild spacetime with line element
\begin{equation}
ds^{2}=-fdt^{2}+f^{-1}dr^{2}+r^{2}d\Omega^{2},
\end{equation}
where $f \equiv 1-\frac{2M}{r}$ and $d \Omega^{2} \equiv d\theta^{2} +\sin^{2}\theta d\phi^{2}$.
The spacetime admits two Killing vectors, $\xi_{(t)}^{\mu}=\partial_t$ and $\xi_{(\phi)}^{\mu}=\partial_\phi$, which in turn give rise to two constants of motion preserved by the MPD equations. In terms of a generic Killing vector $\xi^\mu$, the corresponding constant of motion is given by
\begin{equation}
\label{eqn:COMs}
\Xi=u^{\alpha} \xi_{\alpha}+\frac{\mu}{2} \tilde{S}^{\alpha \beta} \nabla_{\alpha} \xi_{\beta}.
\end{equation}
If $\xi^\mu=\xi_{(t)}^{\mu}$, then $\Xi$ is (minus) the particle's specific energy; if $\xi^\mu=\xi_{(\phi)}^{\mu}$, then it is the particle's angular momentum.

When the secondary's spin vector is aligned (or anti-aligned) with its orbital angular momentum, the MPD equations admit a solution for circular equatorial orbits. Specialising to that case, the orbit is described by
\begin{equation}
\label{eqn:circorbs}
r=r_{p}, \quad \theta=\frac{\pi}{2}, \quad \phi = \Omega t, 
\end{equation}
where $\Omega = \frac{d\phi}{dt}$ is the constant orbital frequency and where the radial and polar components of the secondary's four-velocity vanish, $u^{r}=0 = u^{\theta}$.

For a non-spinning test body in circular, equatorial geodesic motion, the orbital radius and frequency are related by
\begin{equation}
\label{eqn:circorbsNonSpin}
r_p=r_{0}, \quad {\Omega} = \Omega_0 \equiv \sqrt{\frac{M}{r_0^3}}.
\end{equation}

The MPD spin force accelerates the orbit, altering this relationship. To determine the change, one can fix the orbital radius and examine how the spin alters the orbital frequency, or one can fix the frequency and examine how the spin alters the radius.
Previous numerical self-force calculations which included the spin of the secondary parametrised the test-body motion with a fixed-radius parametrisation~ \cite{Han:2010tp, Harms:2015ixa, Akcay:2019bvk, Piovano:2020zin, Piovano:2021iwv}
\begin{equation}
\label{eqn:paramFixedRad}
r_p=r_{0}, \quad \Omega(r_p,\e) ={\Omega}_0+\e \Omega_\chi(r_p) +\mathcal{O}(\e^2).
\end{equation}
However, as illustrated in Sec.~\ref{sec:rw}, it is better to instead adopt a fixed-frequency parametrisation,
\begin{equation}
\label{eqn:paramFixedFreq}
r_p(\Omega,\e)=r_{\Omega}+\e r_\chi(\Omega) +\mathcal{O}(\e^2), \quad \Omega ={\Omega}_0.
\end{equation}
To emphasize that $r_0=r_0(\Omega)$ in Eq.~\eqref{eqn:paramFixedFreq}, we have labelled it as $r_\Omega$. Throughout this paper we work exclusively with the fixed-frequency parametrisation, allowing us to refer simply to the physical frequency $\Omega$ without a 0 adornment. (The notable expception is Appendix \ref{sec:linearisenumerics}, where we discuss the alternative fixed-radius parametrisation.)

To derive the fixed-frequency parametrisation from the equations of motion~\eqref{eqn:MPD} and \eqref{eqn:MPD2} (with hats removed), we take advantage of the fact that we are only looking for solutions valid through order $\e$ (i.e., linear order in the spin force). 
We hence seek a solution of the form 
\begin{equation}
\label{eqn:WorldlineExpand}
z^\alpha(t,\e)=z^\alpha_0(t)+\e z^\alpha_1+\mathcal{O}(\e^2),
\end{equation}
where $z^\mu_0=\big(t,r_\Omega, \pi/2, \Omega t\big)$ and $z^\alpha_1 = r_\chi \delta^\alpha_r$. 

Substituting this ansatz into \eqref{eqn:MPD} and solving order by order in $\e$, we obtain the relations
\begin{equation}
\label{eqn:paramFixedFreqRad}
r_{\Omega}=\frac{M}{(M \Omega)^{2 / 3}}, \quad r_{\chi}=\frac{-F_{(1,\chi)}^r}{3\left({u}_0^{t}\right)^{2} f_\Omega \Omega^{2}},
\end{equation} 
where $f_\Omega\equiv f(r_\Omega)$, $u_0^{\beta}$ is the zeroth-order four-velocity, and $F_{(1,\chi)}^\alpha = -\frac{\mu}{2} R^{r}{}_{\beta \gamma \delta} u_0^{\beta}\tilde S^{\gamma\delta}\delta^\alpha_r$ is the (purely radial) leading-order spin force. The first equality in~\eqref{eqn:paramFixedFreqRad} is simply a restatement of the geodesic relationship~\eqref{eqn:circorbsNonSpin}.

The non-zero components of the four-velocity are given by the circular-orbit condition, $u^{\phi}=u^{t} \Omega$, along with the normalisation $u^{\alpha}u_{\alpha}=-1$. Conveniently, this implies that when parametrised at fixed frequency the four-velocity of the spinning secondary is equal to the corresponding non-spinning geodesic four-velocity, $u^\alpha = {u}_0^\alpha + \mathcal{O}(\e^2)$. 

Writing Eq.~\eqref{eqn:paramFixedFreqRad} more explicitly requires an explicit form for the spin tensor. In the aligned-spin case, the spin vector is given by $S^{\mu}=S^{\theta}\delta^{\mu}_{\theta}$, and the corresponding scaled spin vector $\tilde{S}^{\mu}$ has only one non-zero component, $\tilde{S}^{\theta}=-\frac{\chi}{r_\Omega}$. Introducing a unit vector $\hat z^\alpha \equiv -\frac{1}{r_\Omega}$ along the zeroth-order worldline, we write
\beq\label{S0}
\tilde S^\mu = \chi \hat z^\mu.
\eeq
The scaled spin tensor $\tilde{S}^{\mu \nu}=-\epsilon^{\mu \nu \alpha \beta}\tilde{S}_{\alpha}u_{\beta}$ then has two independent non-zero components \cite{Akcay:2019bvk},
\begin{equation}
\tilde{S}^{tr}=-\frac{\chi}{r_{\Omega}}u_{\phi}=-\tilde{S}^{rt}, \quad \tilde{S}^{r \phi}=-\frac{\chi}{r_{\Omega}}u_{t}=-\tilde{S}^{\phi r}. 
\end{equation}
and we can evaluate the spin force to get
\beq\label{F1 spin}
F_{(1,\chi)}^\alpha = 3 \mu \chi f_\Omega r_\Omega\Omega^3 ({u}_0^t)^2 \delta^\alpha_r.
\eeq

Given this spin force, we find that in the aligned-spin case with a fixed-frequency parametrisation, we have motion described by
\begin{align}
\label{eqn:worldline}
z^\alpha&=\big(t,r_\Omega-\mu \chi \Omega r_\Omega, \pi/2, \Omega t\big)+\mathcal{O}(\e^2),\\
\label{eqn:finalfourvel}
u^\alpha&=\sqrt{\frac{r_\Omega}{r_\Omega-3M}}\left(1,0, 0, \Omega \right)+\mathcal{O}(\e^2),
\end{align}
where we note that the sole spin dependence is in the radial position of the worldline, and where the contravariant components of the four-velocity are independent of $\chi$. Note that the covariant components $u_\alpha$ are not spin independent as $\chi$ enters via the metric components evaluated on the worldline.

Evaluating Eq.~\eqref{eqn:COMs} for the two conserved quantities associated with the timelike and angular Killing vectors yields the specific energy and specific angular momentum, which are given respectively by
\begin{align}
\label{eqn:energy}
E&=E_0 + \e E_\chi = f(r_\Omega)u^t _0 - \frac{\mu \chi}{M} u^t_0 \left(\frac{M}{r_\Omega}\right)^{5/2},\\
\label{eqn:angularmomentum}
J&=J_0 + \e J_\chi = \sqrt{M r_\Omega} u^t_0 + \frac{\mu \chi}{M} u^t_0 \left(\frac{M}{r_\Omega}\right)(r_\Omega-4M),
\end{align}
omitting ${\cal O}(\e^2)$ terms.

\subsection{Stress-energy tensor}
\label{sec:stressenergycomponents}

The stress-energy tensor for the circular-orbit, aligned-spin case can be written in the form
\begin{subequations}
\label{eq:Tmunu}
\begin{align}
\label{eq:explicitmass}
T_{(\mu)}^{\mu \nu} &= \frac{\mu K_{0}^{\mu \nu}}{r^{2} \sin \theta} \delta_{r} \delta_{\theta} \delta_{\phi},\\
\label{eq:explicitspin}
T_{(\chi)}^{\mu \nu} &= \frac{\chi \mu^2}{r^{2} \sin \theta} [ K_{1}^{\mu \nu} \delta_{r} \delta_{\theta} \delta_{\phi} +K_{2}^{\mu \nu} \delta_{r} \delta_{\theta} \delta_{\phi}' +K_{3}^{\mu \nu} \delta_{r}' \delta_{\theta} \delta_{\phi} ],
\end{align}
\end{subequations}
where we use the shorthand notation $\delta_{r}\equiv \delta (r-r_{p})$, $\delta_{\theta} \equiv \delta (\theta-\frac{\pi}{2})$ and $\delta_{\phi}\equiv \delta (\phi- \phi_p)$. The non-zero components of the constant tensors $K_{0}^{\mu \nu}, K_{1}^{\mu \nu}, K_{2}^{\mu \nu}$ and $K_{3}^{\mu \nu}$ are the same as those defined in Ref.~\cite{Akcay:2019bvk} with the identification $r_p \leftrightarrow r_\Omega$ (along with an additional factor of $M^{-1}$ in our definition of $K_{1}^{\mu \nu}, K_{2}^{\mu \nu}$, and $K_{3}^{\mu \nu}$); for completeness we give their full expressions in Appendix \ref{sec:K}.

When we substitute the expansions $r_p=r_\Omega+\e r_\chi+\mathcal{O}(\e^2)$ and $u^\alpha=u^\alpha_0+\mathcal{O}(\e^2)$ into $T^{\mu\nu}_{(\mu)}$ and $T^{\mu\nu}_{(\chi)}$, we can make the trivial replacements $r_p\to r_\Omega$ in \eqref{eq:explicitspin} (and we have preemptively done so in the expressions in Appendix \ref{sec:K}) but we must keep the subleading term in \eqref{eq:explicitmass}, yielding
\begin{equation}
T_{(\mu)}^{\mu \nu} = \frac{\mu K_{0}^{\mu \nu}}{r^3_\Omega} \delta_{\theta} \delta_{\phi}\left(r_\Omega\delta_{r_\Omega} - 2\e r_\chi \delta_{r_\Omega} - \e r_\chi \delta'_{r_\Omega}\right) + \mathcal{O}(\e^2),
\end{equation} 
where $\delta_{r_\Omega}\equiv\delta(r-r_\Omega)$ and $\delta'_{r_\Omega}\equiv\partial_r\delta(r-r_\Omega)$. The total stress-energy~\eqref{split} then becomes
\beq\label{T Taylor}
T^{\mu\nu} = \e T^{\mu\nu}_{(1)} + \e^2 T^{\mu\nu}_{(2)} + \mathcal{O}(\e^3),
\eeq
where
\begin{subequations}
\begin{align}
  T^{\mu\nu}_{(1)} & = \frac{\mu K_{0}^{\mu \nu}}{r^{2}_\Omega \sin \theta} \delta_{r_\Omega} \delta_{\theta} \delta_{\phi},\\
  \label{T2 Taylor}
  T^{\mu\nu}_{(2)} &= -\frac{\mu r_\chi K_{0}^{\mu \nu}}{r^3_\Omega} \delta_{\theta} \delta_{\phi}\left(2 \delta_{r_\Omega} + \delta'_{r_\Omega}\right) + T_{(\chi)}^{\mu \nu}.
\end{align}
\end{subequations}

\section{Two-timescale expansion}
\label{sec:tt expansion}

When the metric perturbations $h^{(n)}_{\alpha\beta}$ are accounted for, the binary system slowly evolves due to dissipation. In this section we show how the linear effects of the particle's spin can easily be incorporated into the two-timescale evolution scheme of Ref.~\cite{Miller:2020bft}. We closely follow the particular formulation in Appendix A of that reference.

Our method assumes the particle's trajectory, its spin, and the spacetime metric only depend on $t$ through the $t$ dependence of a set of mechanical variables $(\phi_p,{\cal J}_I)$. The slow evolution is captured by the parameters ${\cal J}_I = (\Omega, \chi,\delta M, \delta J)$, which evolve on the radiation-reaction timescale $\sim \Omega/\dot\Omega\sim 1/\e$. (Though we find below that $\chi$ is constant at 1PA order.) Here $(\mu\,\delta M, \mu\,\delta J)$ represent corrections to the central black hole's mass and spin, which evolve due to the flux of energy and angular momentum into the black hole; we pull out an overall factor of $\mu$ to make $(\delta M, \delta J)$ order unity. During the slow evolution, the system is assumed to retain a periodic dependence on the particle's orbital phase $\phi_p$, which evolves on the fast timescale $\sim 1/\Omega$. 

Given that the method requires a choice of time coordinate, it will be convenient to adopt the 3+1 split $x^\mu=(t,x^i)$.

\subsection{Orbit, spin, and metric}

We first consider the particle's orbit. In place of Eq.~\eqref{eqn:WorldlineExpand}, we write the coordinate trajectory as $z^\mu(t,\e) = (t,z^i(t,\e))$ and assume that $z^i(t,\e)=z^i(\phi_p(t,\e),{\cal J}_I(t,\e),\e)$. Expanding in powers of $\e$ at fixed $(\phi_p,{\cal J}_I)$, we write
\beq\label{eqn: tt worldline}
z^i(\phi_p,{\cal J}_I,\e) = z_0^i(\phi_p,\Omega)+\e z_1^i({\cal J}_I)+O(\e^2),
\eeq
where the leading-order trajectory is
\beq
z_0^i(\phi_p,\Omega) = (r_0(\Omega),\pi/2,\phi_p),
\eeq
and the subleading term is a purely radial correction
\beq
z_1^i({\cal J}_I) = (r_1({\cal J}_I),0,0).
\eeq
We continue to define the frequency as
\beq\label{Omegadef}
 \frac{d\phi_p}{dt} \equiv \Omega.
\eeq

The above ansatz represents an orbit that remains in the equatorial plane, with a slowly evolving radius and frequency. Accordingly, we seek a solution in which the spin remains orthogonal to the equatorial plane,
\beq\label{tt spin}
\tilde S^\alpha({\cal J}_I) = \chi \hat z^\alpha(\Omega)+O(\e),
\eeq
in analogy with \eqref{S0}.

Following the same pattern, we expand the metric as\footnote{$\e^2\ln\e$ terms also appear, at least in the Lorenz and similar gauges. We hide that logarithmic dependence in $h^{(2)}_{\mu\nu}$. As discussed in Ref.~\cite{Miller:2020bft}, it is also preferable to adopt a hyperboloidal time coordinate $s= t - k(r^*)$. We elide that detail here, as it is not important for the spin contributions at 1PA order.} 
\begin{multline}
\gexact_{\mu\nu} = g_{\mu\nu}(x^i) + \e h^{(1)}_{\mu\nu}(x^i,\phi_p,\Omega,\delta M, \delta J)\\ + \e^2 h^{(2)}_{\mu\nu}(x^i,\phi_p,{\cal J}_I) + O(\e^3),\label{g tt expansion}
\end{multline}
with the assumption that each term is periodic in $\phi_p=\phi_p(t,\e)$. For simplicity, we suppress dependence on $M$ and $\mu$, with the understanding that $h^{(n)}_{\mu\nu}$ is equal to $\mu^n$ times a $\mu$-independent function of $(x^i,\phi_p,{\cal J}_I)$.

The first-order perturbation $h^{(1)}_{\mu\nu}(x^i,\phi_p,\Omega,\delta M, \delta J)$ is linear in $\delta M$ and $\delta J$, and it will be convenient to peel off that dependence, writing
\begin{multline}
h^{(1)}_{\mu\nu} = h^{(1)}_{\mu\nu}(x^i,\phi_p,\Omega, 0, 0) + \delta M\, h^{(\delta M)}_{\mu\nu}(x^i) \\+ \delta J\, h^{(\delta J)}_{\mu\nu}(x^i).
\end{multline}
If we replaced $\phi_p$ with its geodesic expression $\Omega t$, then the first term would be the standard perturbation due to a point mass on a circular geodesic with frequency $\Omega$. The terms $\delta M\, h^{(\delta M)}_{\mu\nu}(x^i)$ and $\delta J\, h^{(\delta J)}_{\mu\nu}(x^i)$ are linear perturbations toward a Kerr black hole with mass $M+\mu\,\delta M$ and angular momentum $\mu\,\delta J$. We show below that at 1PA order, $\delta M$ and $\delta J$ do not couple to the spin $\chi$.

When substituting these expansions into the equations of motion and field equations, we apply the chain rule 
\beq
\frac{\partial}{\partial x^\alpha} = e^i_\alpha \frac{\partial}{\partial x^i} + t_\alpha\left(\frac{d\phi_p}{dt}\frac{\partial}{\partial\phi_p} +\frac{d{\cal J}_I}{dt}\frac{\partial}{\partial {\cal J}_I}\right),
\eeq
where $e^i_\alpha \equiv \frac{\partial x^i}{\partial x^\alpha}$ and $t_\alpha \equiv \partial_\alpha t$. This implies the expansion
\beq\label{nabla tt}
\nabla_\alpha = \nabla^{(0)}_\alpha + \e t_\alpha \vec{\partial}_{\cal V} + {\cal O}(\e^2),
\eeq
where the zeroth-order covariant derivative is 
\beq
\nabla^{(0)}_\alpha = e^i_\alpha\frac{\partial}{\partial x^i} + t_\alpha \Omega\frac{\partial}{\partial\phi_p} + \text{Christoffel terms}.\label{nabla0}
\eeq
${\cal V}_I=(F^{(0)}_\Omega, F^{(1)}_\chi, F^{(1)}_{\delta M}, F^{(1)}_{\delta J})$ is the leading-order velocity through parameter space, and
\begin{align}
\vec{\partial}_{\cal V}&\equiv{\cal V}_I\frac{\partial}{\partial J_I} \\
&= F_\Omega^{(0)}\frac{\partial}{\partial\Omega} + F_\chi^{(1)}\frac{\partial}{\partial\chi} + F_{\delta M}^{(1)}\frac{\partial}{\partial\delta M} + F_{\delta J}^{(1)}\frac{\partial}{\partial\delta J}.
\end{align}
$\nabla^{(0)}_\alpha$ acts at fixed parameter values; its action on $h^{(1)}_{\alpha\beta}(x^i,\phi_p,\Omega, 0, 0)$ is identical to the action of $\nabla_\alpha$ on the linear metric perturbation from a point mass on a circular geodesic. The directional derivative $\vec{\partial}_{\cal V}$ then accounts for the system's slow movement through the parameter space.

\subsection{Evolution equations}

The rates of change of the parameters ${\cal J}_I$ are likewise expanded in powers of $\e$ at fixed $(\phi_p,{\cal J}_I)$:
\begin{align}
    \frac{d\Omega}{dt} &= \e F_{\Omega}^{(0)}(\Omega) + \e^2 F_\Omega^{(1)}({\cal J}_I) + {\cal O}(\e^3),\label{Omegadot}\\        
    \frac{d\chi}{dt} &= \e F_\chi^{(1)}({\cal J}_I) + {\cal O}(\e^2),\label{chidot}\\
    \frac{d\delta M}{dt} &= \e F_{\delta M}^{(1)}(\Omega) + {\cal O}(\e^2),\label{Mdot}\\
    \frac{d\delta J}{dt} &= \e F_{\delta J}^{(1)}(\Omega) + {\cal O}(\e^2).\label{Jdot}
\end{align}
From these expansions we obtain the expansion for the coordinate velocity, 
\beq
\dot z^\alpha \equiv \frac{dz^\alpha}{dt} = \dot z^\alpha_0(\Omega) + \e \dot z^\alpha_1(\Omega) +O(\e^2),
\eeq
where
\beq
\dot z_0^\alpha \equiv \frac{dz_0^\alpha}{dt} = (1,0,0,\Omega),
\eeq
and
\beq\label{z1dot}
\dot z_1^\alpha = (0,\dot r_0,0,0),
\eeq
with $\dot r_0(\Omega)=\frac{dr_0}{d\Omega}F_{\Omega}^{(0)}$; the proper four-velocity is $u^\alpha=u^\alpha_0(\Omega)+\e u^t_0(\Omega)\dot z^\alpha_1(\Omega)+\mathcal{O}(\e^2)$, with $u^\alpha_0=u^t_0\dot z^\alpha_0$ as in~\eqref{eqn:finalfourvel}. The driving forces $F_{Y}^{(n)}$, which govern the evolution, are to be determined from the equations of motion~\eqref{selfforceEOMs} and from the Einstein field equations.

In the above expansions, the numerical labels within parentheses denote the post-adiabatic order at which the quantity enters, while the numeric labels without parentheses correspond to the explicit powers of $\e$.\footnote{This statement assumes that we calculate $F^{(0)}_{\Omega}$ using the local self-force. If we instead calculate it from energy fluxes to infinity and into the horizon, then the leading horizon fluxes $F_{\delta M}^{(1)}(\Omega)$ and $F_{\delta J}^{(1)}(\Omega)$ enter at 0PA order, and the first subleading horizon fluxes enter at 1PA order. However, $\delta M$ and $\delta J$ themselves only enter at 1PA order in either approach.} 
We have foreshadowed the structure of the solution by indicating that 0PA quantities only depend on $(\phi_p,\Omega)$, and that $d{\cal J}_I/dt$ is independent of $\phi_p$ (an essential requirement for the separation between slow and fast evolution). 

To expand the equations of motion~\eqref{selfforceEOMs}, we also expand the forces and torque as
\begin{align}
F^\alpha &= \e F^\alpha_1({\cal J}_I) + \e^2 F^\alpha_2({\cal J}_I)+{\cal O}(\e^3),\\
N^\alpha &= \e N^\alpha_1({\cal J}_I) +{\cal O}(\e^2).
\end{align}
Here the numeric labels correspond to the explicit powers of $\e$, following the usual nomenclature for ``first-order'' and ``second-order'' self-forces. The spin-dependent contributions are given by the expansions of~\eqref{spin F and N}. More precisely, the first-order spin force $F^\alpha_{(1,\chi)}(\Omega,\chi)$ is given by the test-body force~\eqref{F1 spin}; $F^\alpha_{(2,\chi)}({\cal J}_I)$ is given by the remainder of Eq.~\eqref{spin force}; and $N^\alpha_1$ is given by Eq.~\eqref{spin torque} evaluated at  $r_p=r_\Omega$ and $u^\alpha=u^\alpha_0$. We will not require a more explicit expression for $F^\alpha_{(2,\chi)}$.
The torque reduces to
\beq\label{1PA torque}
N^\mu_1 = -\frac{\chi}{2}\hat z^\mu(g^{\alpha\beta}-\hat z^\alpha \hat z^\beta)u_0^\gamma \nabla^0_\gamma h^{\mathrm{R}(1)}_{\alpha\beta} = 0. 
\eeq
To see why this vanishes, note that it can be written as $-\frac{\chi}{2}\hat z^\mu u_0^\gamma \nabla^0_\gamma\left[(g^{\alpha\beta}-\hat z^\alpha \hat z^\beta)h^{\mathrm{R}(1)}_{\alpha\beta}\right]$. For our quasicircular, spin-aligned system, the quantity in square brackets is constant along the zeroth-order worldline $z^\mu_0$ (at fixed $\Omega$). The derivative therefore vanishes.

Substituting all of the above expansions into Eqs.~\eqref{selfforceorbit} and~\eqref{selfforcespin}, we can straightforwardly solve order by order in $\e$, equating coefficients of powers of $\e$ at fixed ${\cal J}_I$ rather than at fixed $t$. We obtain 
\begin{align}\label{radius}
r_0(\Omega) = r_\Omega, \quad r_1({\cal J}_I) = -\frac{F_1^r({\cal J}_I)}{3(u_0^t)^2 f_\Omega\Omega^2}
\end{align}
from the conservative sector [the radial component of Eq.~\eqref{selfforceorbit}], in perfect analogy with Eq.~\eqref{eqn:paramFixedFreqRad}. From the dissipative sector [Eq.~\eqref{selfforcespin} and the $t$ or $\phi$ component of~\eqref{selfforceorbit}], we obtain
\begin{align}
F_\Omega^{(0)} &= -\frac{3 f_\Omega\Omega F^t_1(\Omega)}{y(u_0^t)^4(1-6y)},\label{Omegadot0}\\
F_\Omega^{(1)} &= -\frac{3 f_\Omega\Omega F^t_2({\cal J}_I)}{y(u_0^t)^4(1-6y)} 
 - \frac{2 \vec{\partial}_{\cal V}F_1^r({\cal J}_I)}{\sqrt{y}(u_0^t)^4 f_\Omega(1-6y)}\nonumber\\
&\quad - \frac{4(1-6y+12y^2)F_1^r({\cal J}_I) F_1^t(\Omega)}{y^{3/2} (u_0^t)^6f_\Omega(1-6y)^2}\label{Omegadot1},\\
F^{(1)}_\chi &= N^z_1 = 0.\label{chidot1}
\end{align}
Here 
\beq
y\equiv M/r_\Omega = (M\Omega)^{2/3}.
\eeq
Equation~\eqref{chidot1} shows that the spin magnitude is constant at 1PA order. 

The equations of motion do not determine the driving forces $F_{\delta M}^{(1)}(\Omega)$ and $F_{\delta J}^{(1)}(\Omega)$. However, as shown in Ref.~\cite{Miller:2020bft}, the second-order Einstein equations [\eqref{tt EFE2} below] dictate that these are the usual fluxes of energy and angular momentum through the horizon due to a point mass on a circular geodesic orbit of frequency $\Omega$  (reviewed in Sec.~\ref{sec:flux} below).

It is easy to see that if we rewrite Eqs.~\eqref{Omegadef} and \eqref{Omegadot}--\eqref{Jdot} in terms of  a ``slow time'' variable $\tilde t\equiv\e t$, then the equations have asymptotic solutions
\begin{align}
    \phi_p &= \e^{-1}\phi^{(0)}_p(\tilde t)+\phi^{(1)}_p(\tilde t)+{\cal O}(\e),\label{phip - slow t}\\
    \Omega &= \Omega^{(0)}(\tilde t)+\e\,\Omega^{(1)}(\tilde t)+{\cal O}(\e^2),\\
    \chi &= \chi^{(1)} +{\cal O}(\e),\\
    \delta M &= M^{(1)}(\tilde t)+{\cal O}(\e),\\
    \delta J &= J^{(1)}(\tilde t)+{\cal O}(\e),\label{dJ - slow t}
\end{align}
with constant $\chi^{(1)}$, with $d\phi^{(n)}_p/d\tilde t=\Omega^{(n)}(\tilde t)$, and with easily worked out equations for $d\Omega^{(n)}/d\tilde t$. These expansions  in powers of $\e$ at fixed slow time make clear the structure of the solution on the radiation-reaction timescale $t\sim M/\e$ ($\tilde t\sim M$). However, they are primarily useful at the final waveform-generation stage, where they allow one to solve for the coefficients $\phi^{(n)}_p$, $\Omega^{(n)}$, $M^{(1)}$, and $J^{(1)}$  without specifying a value of the mass ratio. Prior to that stage, we treat $(\phi_p,\Omega,\chi,\delta M, \delta J)$ as independent variables on the binary's phase space.

\subsection{Stress-energy tensor}

Substituting the expansions~\eqref{eqn: tt worldline}--\eqref{z1dot} into the stress-energy~\eqref{split}, we obtain
\begin{align}
\!\!T^{\mu\nu} = \e T_{(1)}^{\mu\nu}(x^i,\phi_p,\Omega) + \e^2 T_{(2)}^{\mu\nu}(x^i,\phi_p,{\cal J}_I) + O(\e^3).\!\label{T tt expansion}
\end{align} 
The leading term is still given by the leading term in Eq.~\eqref{T Taylor} except that in $\delta_\phi$ we do not replace $\phi_p$ with $\Omega t$. The subleading term is now
\begin{multline}
\!\!\!\!\!T_{(2)}^{\alpha\beta} =\frac{u_0^t}{r_0^3}\left[2\left(r_0\dot z^{(\alpha}_0\dot z_1^{\beta)}-\dot z^\alpha_0 \dot z^\beta_0 r_1\right)\delta_r - \dot z_0^\alpha \dot z_0^\beta r_0 r_1\delta'_r\right]\delta_\theta\delta_\phi \\
+ T_{(\chi)}^{\alpha\beta},\label{T2 multiscale}
\end{multline}
which is identical to Eq.~\eqref{T2 Taylor} except for (i) the change $r_\chi\to r_1$, (ii) the addition of the $\dot z_1^\beta$ term, and (iii) the fact that we again do not replace $\phi_p$ with $\Omega t$. In this expression, $r_1$ receives a contribution from the first-order self-force as well as from the first-order spin force~\eqref{F1 spin}. Since the spin does not contribute to $\dot z_1^{\beta}$, its total contribution to $T^{\alpha\beta}_{(2)}$ therefore remains precisely~\eqref{T2 Taylor}.

\subsection{Field equations and Fourier expansions}

Substituting the expansions~\eqref{g tt expansion} and \eqref{T tt expansion} into the Einstein equations $G_{\mu\nu}[\gexact]=T_{\mu\nu}$ and equating coefficients of powers of $\e$ at fixed $(\phi_p,{\cal J}_I)$, we obtain the hierarchy 
\begin{subequations}
\label{tt EFE}
\begin{align}
 G_{\mu\nu}[g] &= 0,\\
 G^{(1,0)}_{\mu\nu}[h^{(1)}] &= 8\pi T^{(1)}_{\mu\nu},\label{tt EFE1}\\
 G^{(1,0)}_{\mu\nu}[h^{(2)}] &= 8\pi T^{(2)}_{\mu\nu} - G^{(2,0)}_{\mu\nu}[h^{(1)},h^{(1)}] \nonumber\\
 &\quad -G^{(1,1)}_{\mu\nu}[h^{(1)}].\label{tt EFE2}
\end{align}
\end{subequations}
The operators $G^{(n,j)}_{\mu\nu}$ act on functions of $(x^i,\phi_p,{\cal J}_I)$. If we first expand $G_{\mu\nu}[g+h]$ in powers of $h_{\mu\nu}$, as $G_{\mu\nu}+G^{(1)}_{\mu\nu}[h]+G^{(2)}_{\mu\nu}[h,h]+\ldots$, then $G^{(n,j)}_{\mu\nu}$ is derived from $G^{(n)}_{\mu\nu}$ using the expansion of $\nabla_\alpha$ in Eq.~\eqref{nabla tt}. Using that expansion, we see that $G^{(1,0)}_{\mu\nu}$ is the standard linearized Einstein tensor with $\nabla_\alpha\to \nabla^{(0)}_\alpha$. $G^{(n,1)}_{\mu\nu}$ is given by the terms in $G^{(n)}_{\mu\nu}$ that are linear in the velocity ${\cal V}_I$; since these do not couple to spin terms at 1PA order, we will not display them here, but they can be extracted from Ref.~\cite{Miller:2020bft}. 

The linear-in-spin equations are
\beq\label{tt spin EFE}
G^{(1,0)}_{\mu\nu}[h^{(2,\chi)}] = 8\pi T^{(2,\chi)}_{\mu\nu},
\eeq
where $T^{(2,\chi)}_{\mu\nu}$ is given by Eq.~\eqref{T2 Taylor} as described around Eq.~\eqref{T2 multiscale}. Note that $h^{(2,\chi)}_{\mu\nu}$, and the associated regular field $h^{\mathrm{R}(2,\chi)}_{\mu\nu}$, includes the effect of $r_\chi$ in addition to the effect of $T^{\mu\nu}_{(\chi)}$. This means that $h^{\mathrm{R}(2,\chi)}_{\mu\nu}$ differs from the field $h^{\mathrm{R}(\chi)}_{\mu\nu}$ in Sec.~\ref{sec:gsf} by terms proportional to $r_\chi$ (plus order-$\e^3$ differences).

Since all functions of $\phi_p$ are periodic, we can expand them in Fourier series. For example
\beq\label{two-time modes}
h^{(n)}_{\alpha\beta}(x^i,\phi_p,{\cal J}_I) = \sum_{m=-\infty}^\infty h^{(n,m)}_{\alpha\beta}(x^i,{\cal J}_I)e^{-im\phi_p}.
\eeq
We then have $\frac{\partial}{\partial\phi_p} \to -im$ when acting on individual modes, implying
\beq\label{del0 modes}
\nabla^{(0)}_\alpha \to e^i_\alpha\frac{\partial}{\partial x^i} -i t_\alpha \omega_m + \text{Christoffel terms},
\eeq
where $\omega_m \equiv m\Omega$. The label $m$ here also serves as the azimuthal mode number, such that
\beq
h^{(n)}_{\alpha\beta}(x^i,\phi_p,{\cal J}_I) = \sum_{m=-\infty}^\infty h^{(n,m)}_{\alpha\beta}(r,\theta,{\cal J}_I)e^{im(\phi-\phi_p)}.
\eeq
(We abuse notation by using $h^{(n,m)}_{\alpha\beta}$ for the coefficients in both decompositions.)

The action of $\nabla^{(0)}_\alpha$ in Eq.~\eqref{del0 modes} is identical to the action of an ordinary covariant derivative acting on a Fourier series $\sum_{m=-\infty}^\infty h^{(n,m)}_{\alpha\beta}(x^i)e^{-im\Omega t}$, even though $\phi_p$ in Eq.~\eqref{two-time modes} is {\em not} equal to $\Omega t$. Analogously, when acting on a Fourier expansion of the form~\eqref{two-time modes}, the leading-order linearized Einstein tensor $G^{(1,0)}_{\mu\nu}$ is identical to the ordinary linearized Einstein tensor acting on an ordinary Fourier series with modes $e^{-im\Omega t}$. Equations~\eqref{tt EFE1} and \eqref{tt EFE2} therefore reduce to the familiar form of linearized Einstein equations in the frequency domain.

\subsection{Summary: two-timescale evolution with spin}
\label{sec:SummaryTT}

Our analysis has shown that linear-in-spin effects are easily incorporated into the two-timescale evolution and waveform-generation scheme of Ref.~\cite{Miller:2020bft}. That scheme can be summarized in two conceptually simple steps:\footnote{There are also two simplifications that are important in practice but not conceptually essential. First, the offline computations are done on the one-dimensional space of $\Omega$ values rather than the three-dimensional space of $(\Omega,\delta M,\delta J)$ values because we only require effects linear in $\delta M$ and $\delta J$, allowing us to compute the coefficients of those effects without specifying values of $(\delta M,\delta J)$. Second, by assuming the ansatzes~\eqref{phip - slow t}--\eqref{dJ - slow t}, we can convert Eqs.~\eqref{Omegadef} and \eqref{Omegadot}--\eqref{Jdot} into equations for the coefficients $\phi^{(0)}_p(\tilde t)$, $\phi^{(1)}_p(\tilde t)$, etc., which can be solved for and stored without specifying $\mu$. The inspiral trajectory and waveform for a given $\mu$ can then be generated effectively instantly using the stored solutions.}
\begin{enumerate}
\item {\em Offline computations}. On a grid of $\Omega$ values, solve the field equations~\eqref{tt EFE1} and~\eqref{tt EFE2} for the mode amplitudes $h^{(n,m)}_{\alpha\beta}$. From them, compute the forcing functions $F^{(0)}_\Omega(\Omega)$, $F^{(0)}_{\delta M}(\Omega)$, $F^{(0)}_{\delta J}(\Omega)$, and $F^{(1)}_\Omega(\Omega,\chi=0,\delta M,\delta J)$. These calculations only involve the coefficients of the phase factors $e^{-im\phi_p}$, never the orbital phase $\phi_p(t,\e)$ itself; the orbital phase factors out of the computations.
\item {\em Online simulation}. Using the stored forcing functions, choose a value of $\mu$ and solve Eqs.~\eqref{Omegadef} and \eqref{Omegadot}--\eqref{Jdot} for the phase-space trajectory $$(\phi_p(t,\e),\Omega(t,\e),\chi=0,\delta M(t,\e), \delta J(t,\e)).$$
From that trajectory and the mode amplitudes, $h^{(n,m)}_{\alpha\beta}$, generate the waveform $\lim_{r\to\infty}r\sum_{n,m} \e^n h^{(n,m)}_{\alpha\beta}(x^i,{\cal J}_I(t,\e))e^{-im\phi_p(t,\e)}$.
\end{enumerate}

To incorporate the spin into this framework, we simply add the following to the first step:
\begin{itemize}
\item[1.*] On the grid of $\Omega$ values, solve the field equation~\eqref{tt spin EFE} for the linear-in-spin contribution to the mode amplitudes $h^{(2,m)}_{\alpha\beta}$. From them and $h^{(1,m)}_{\alpha\beta}$, compute the linear-in-spin contribution to $F^{(1)}_{\Omega}$.
\end{itemize}
This can be done without specifying a value of $\chi$. In Step 2, we then simply set a freely specifiable nonzero value $\chi$, which remains constant by virtue of Eq.~\eqref{chidot1}, add the linear-in-spin term to $F^{(1)}_{\Omega}$ and $h^{(2,m)}_{\alpha\beta}$, and proceed as above to generate the waveform. 

In principle, the spin contribution to $F^{(1)}_{\Omega}$ can be computed directly from the local expression~\eqref{Omegadot1}. However, we can also extract it from the asymptotic fluxes of gravitational waves. As shown in~\cite{Akcay:2019bvk}, the quantity $\Xi$ defined in Eq.~\eqref{eqn:COMs} (the energy or angular momentum) satisfies a flux-balance law: neglecting higher-order spin effects, and neglecting all other ${\cal O}(\e^2)$ terms in the equation of motion, the rate of change of $\Xi$ due to the local force and torque, which can be written as
\begin{equation}\label{<dXidtau>}
\left\langle\frac{D \Xi}{d \tau}\right\rangle=\frac{1}{2}\left\langle u^{\alpha} u^{\beta} \mathcal{L}_{\xi} h_{\alpha \beta}^{\rm R}+\mu \tilde{S}^{\alpha \beta} u^{\gamma} \nabla_{\alpha} \mathcal{L}_{\xi} h_{\beta\gamma}^{\rm R}\right\rangle,
\end{equation}
is equal (with opposite sign and up to a factor of $u^t$) to the total flux of gravitational-wave energy (or angular momentum) out to infinity and into the black hole.\footnote{The balance law holds for a generic bound orbit in Kerr spacetime, in which case this statement applies on average, with the average over radial and polar oscillations denoted by angle brackets. For quasicircular, equatorial orbits, no average is required.}
For example, with the caveats of what is being neglected, the rate of change of the energy $E$ is
\begin{equation}\label{balance law}
\frac{dE}{d t}=-(\dot{E}^{+} + \dot{E}^{-})\equiv - {\cal F},
\end{equation} 
where $\dot{E}^{+}$ and $\dot{E}^{-}$ are the gravitational-wave energy fluxes (per unit $\mu$) to infinity and into the horizon, respectively. 

The energy has the form $E({\cal J}_I)=E_0(\Omega)+\e E_\chi(\Omega,\chi) + \e E_{SF}(\Omega,\delta M,\delta J)+{\cal O}(\e^2)$, where the first two terms are given by the test-body energy~\eqref{eqn:energy}, and the final term comes from the first-order radial self-force term in $-u^\alpha\xi_\alpha = f(r_p)u^t$. Substituting this expression for $E$ into the balance law~\eqref{balance law}, applying the chain rule, substituting $d\Omega/dt = \e F^{(0)}_{\Omega}+{\cal O}(\e^2)$ and $E_0=f_\Omega u^t_0(\Omega)$, and rearranging for $F^{(0)}_{\Omega}$, we immediately find
\beq
\label{eq:FOmega0}
F^{(0)}_{\Omega}(\Omega) =  \frac{3\Omega}{y (u^t_0)^3(1-6y)}{\cal F}_{(1)}(\Omega).
\eeq
This is the standard adiabatic, flux-driven evolution of the frequency, and ${\cal F}_{(1)}(\Omega)$ is identical to the standard leading-order flux due to a point mass on a geodesic circular orbit of frequency $\Omega$.

We can similarly pick off the linear-in-spin, ${\cal O}(\e^2)$ terms in the balance law~\eqref{balance law}. We write their contribution as
\beq\label{balance law linear}
\left(\frac{dE}{dt}\right)_{\chi} = -{\cal F}_{\chi},
\eeq
noting that these denote coefficients of $\chi$ rather than including the linear factor of $\chi$. Since we are not neglecting any  linear-in-spin, ${\cal O}(\e^2)$ terms, this formula is exact, unlike~\eqref{balance law}.  Again applying the chain rule, using $d\chi/dt={\cal O}(\e^2)$, and picking out the linear-in-spin terms, we find the left-hand side of Eq.~\eqref{balance law linear} evaluates to
\beq
\left(\frac{dE_{\chi}}{dt}\right)_{\chi} = \frac{\partial E_0}{\partial\Omega}F^{(1)}_\Omega+ \left(\frac{\partial E_{\chi}}{\partial\Omega}\right)_\chi F^{(0)}_\Omega.
\eeq
Substituting this into Eq.~\eqref{balance law linear} and rearranging, we obtain our desired result:
\begin{multline}
F^{(1,\chi)}_\Omega(\Omega) = \frac{3\Omega}{y (u^t_0)^3(1-6y)}{\cal F}_{\chi}(\Omega) \\
- \frac{3\mu\Omega^2(5-12y)}{(u^t_0)^3 y(1-6y)^2} {\cal F}_{(1)}(\Omega). \label{F1spin}
\end{multline}
This is the most essential input for a 1PA evolution including spin. Its main ingredient, ${\cal F}_{\chi}(\Omega)$, has the distinct advantage of being more easily computatble than the local metric perturbation, as it is derived from the amplitudes $h^{(n,m)}_{\alpha\beta}$ at infinity and the black hole's horizon. It has been calculated previously by several groups using a fixed-radius formulation \cite{Akcay:2019bvk, Piovano:2020zin, Piovano:2021iwv}; we review its computation using a more efficient fixed-frequency formulation in the next section. 

Before moving to the next section, we note that although we have not included quadratic-in-spin terms in our calculations, they would not enter into the 1PA waveform. This follows from the fact that at ${\cal O}(\e^2)$, they only enter in the form of test-body terms in the background spacetime. Such terms are known to be purely conservative~\cite{VojtechChat, Vines:2016unv}, and only dissipative ${\cal O}(\e^2)$ terms can enter at 1PA (specifically, through $F_\Omega^{(1)}$).

\section{Regge-Wheeler-Zerilli formalism}
\label{sec:rw}

The Regge-Wheeler-Zerilli (RWZ) formalism \cite{Regge1957, Zerilli1970} specialises the linearized Einstein field equations to perturbations of a Schwarzschild background spacetime. The RWZ equations follow from a tensor spherical harmonic decomposition of the field equations which separates the odd and even parity perturbations. Physically speaking, the even parity sector describes fields that are invariant under the transformation $(\theta, \phi) \rightarrow(\pi-\theta, \pi+\phi)$ while the odd parity sector describes fields that are changed by a factor of $-1$ under the same transformation.

In the RWZ formalism, two master functions satisfying the RWZ equations are defined in terms of the metric perturbation amplitudes, one for the odd parity sector and one for the even parity sector. The master functions can then be used to reconstruct the metric perturbation amplitudes and thus the metric perturbation. We make a gauge choice in this final step: the odd parity perturbations are fixed into the Regge-Wheeler gauge while the even parity perturbations are fixed into the Zerilli gauge. 

In this work we follow the gauge-invariant approach to the RWZ formalism detailed in \cite{PhysRevD.71.104003,PhysRevD.82.084010}. In the even parity sector we use the Zerilli-Moncrief \cite{Moncrief1974} master function, while in the odd parity sector we use the Cunningham-Price-Moncrief \cite{Cunningham1979} master function.

\subsection{The Regge-Wheeler-Zerilli equation}
\label{sec:rwz_eqn}
Both the Zerilli-Moncrief and the Cunningham-Price-Moncrief master functions satisfy a wave equation (the RWZ equation), which in the frequency domain is of the form
\begin{equation}
\label{eqn:RWZfreq}
\left[\dfrac{\partial^{2}}{\partial r_{*}^{2}} -V_{\ell}(r) + \omega^{2} \right]\psi_{\ell m \omega}(r)=Z_{\ell m \omega}(r),
\end{equation}
where $r_{*}=r+2M\ln(r/2M -1)$ is the Schwarzschild tortoise coordinate such that $\dfrac{d r}{d r_{*}} = f(r)$, and where $\omega=m \Omega$ for circular orbits. The form of the potential $V_{\ell}(r)$ and the source term $Z_{\ell m \omega}$ depends on the master function, $\psi_{\ell m\omega}$, and therefore is different in each parity sector, with the Zerilli potential in the even sector and the Regge-Wheeler potential in the odd sector. In each parity sector, the source term derives from the stress-energy tensor due to a small spinning point particle, Eq.~\eqref{T Taylor}, and is of the form
\begin{equation}
 \label{eqn:explicitsource}
 Z_{\ell m\omega} = \left( \bar{G}_{\ell m\omega}\delta_{r_\Omega} + \bar{F}_{\ell m\omega} \delta_{r_\Omega}' + \bar{H}_{\ell m\omega} \delta_{r_\Omega}'' \right),
\end{equation}
where as previously we use the shorthand $\delta_{r_\Omega}\equiv\delta(r-{r_\Omega})$. The functions $\bar{G}_{\ell m\omega}$, $\bar{F}_{\ell m\omega}$ and $\bar{H}_{\ell m\omega}$ depend only on ${r_\Omega}$, $\chi$, $\mu$ and $M$ \footnote{In the non-spinning limit, $\bar{H}_{\ell m\omega}\rightarrow0$. When writing the source at fixed frequency, in the even parity sector it also turns out that $\bar{H}^{\rm even}_{\ell m\omega}=0.$}; explicit expressions for these and for the potentials $V_\ell(r)$ are given in Appendix \ref{sec:sources}.

\subsection{Retarded solutions to the RWZ equations}
\label{sec:retsoln}

The RWZ equation, Eq.~\eqref{eqn:RWZfreq}, admits two linearly independent homogeneous solutions. There is flexibility in the particular choice of basis of homogeneous solutions. For radiative ($\omega \ne 0$) modes we choose to work with ``in'' and ``up''  solutions, which satisfy boundary conditions representing radiation that is purely ingoing into the future horizon and purely outgoing to future null infinity, respectively.
Using hats to signify that they are unit-normalised homogeneous solutions, the ``in'' and ``up''  solutions therefore have the asymptotic behaviour
\begin{subequations}
\begin{align}
\label{bcs}
 \hat{R}_{\ell m\omega}^-(r \rightarrow 2M) &=e^{-i \omega_m r_*}, \\
 \hat{R}_{\ell m\omega}^+(r \rightarrow \infty) &=e^{i \omega_m r_*}.
\end{align}
\end{subequations}
For the non-radiative ($\omega=0$) modes we use analytic solutions in terms of hypergeometric functions,
\begin{subequations}
\begin{align}
\label{eq:homogeneous bcs}
 \hat{R}_{\ell m0}^-(r) &= \, _2F_1\left(l-s+1,l+s+1;1;f(r)\right)\\
 \hat{R}_{\ell m0}^+(r) &= \, _2F_1\left(l-s+1,l+s+1;2 (l+1);2M/r\right),
\end{align}
\end{subequations}
where $s=2$ for the gravitational RWZ equations.
Here, we adopt the same notation as for the radiative modes but our solutions are now chosen on the basis that they are regular at the horizon and at infinity, respectively.

In this work we make use of the \texttt{ReggeWheeler} package from the Black Hole Perturbation Toolkit \cite{BHPToolkit} to obtain numerical solutions to the homogeneous RWZ equations. We then use the method of variation of parameters to find the inhomogeneous solution of \eqref{eqn:RWZfreq} in terms of these homogeneous solutions,
\begin{equation}
\label{eqn:varofparams}
\psi_{\ell m\omega}(r) = c^+_{\ell m\omega}(r)\hat{R}_{\ell m\omega}^+(r) + c^-_{\ell m\omega}(r)\hat{R}_{\ell m\omega}^-(r),
\end{equation}
where
\begin{subequations}
\begin{align}
\label{eqn:match1}
c^+_{\ell m\omega}(r)\equiv \frac{1}{W_{\ell m\omega}}\int^r \frac{\hat{R}_{\ell m\omega}^-(r')  Z_{\ell m\omega}(r')}{f(r')}dr', \\
\label{eqn:match2}
c^-_{\ell m\omega}(r)\equiv \frac{1}{W_{\ell m\omega}}\int_{r} \frac{\hat{R}_{\ell m\omega}^+(r')  Z_{\ell m\omega}(r')}{f(r')}dr' ,
\end{align}
\end{subequations}
and the factor of $f(r)$ in the denominator of the integrand comes from changing integration variable from $r_{*}$ to $r$. In the method of variation of parameters, the Wronskian
\begin{equation}
W_{\ell m\omega}\equiv \hat{R}_{\ell m\omega}^-\frac{d\hat{R}_{\ell m\omega}^+}{dr_*}-\hat{R}_{\ell m\omega}^+\frac{d\hat{R}_{\ell m\omega}^-}{dr_*},
\end{equation}
typically appears inside the integral. However, since there is no first-derivative term in Eq.~\eqref{eqn:RWZfreq}, by Abel's identity the Wronskian is a constant so it may be taken outside of the integral and evaluated at any convenient radius.

Substituting the source, $Z_{\ell m\omega}(r)$, given by Eq.~\eqref{eqn:explicitsource} into Eqs.~\eqref{eqn:match1} and \eqref{eqn:match2}, integrating by parts and paying careful attention to the boundary terms, the inhomogeneous solution in Eq.~\eqref{eqn:varofparams} becomes
\begin{equation}
\label{eqn:formR}
\psi_{\ell m\omega}(r) = R_{\ell m\omega}^+(r)\Theta^+_{r_\Omega} + R_{\ell m\omega}^-(r)\Theta^-_{r_\Omega} + X_{\ell m\omega}\delta_{r_\Omega}, 
\end{equation} 
where we have denoted the Heaviside step functions by $\Theta^+_{r_\Omega}\equiv \Theta[r-r_\Omega]$ and $\Theta^-_{r_\Omega}\equiv \Theta[r_\Omega-r]$, and  $X_{\ell m\omega}\equiv \frac{\bar{H}_{\ell m\omega}}{f^2_{\Omega}}$ is a constant. In the above expressions we have introduced the shorthand $R^\pm_{\ell m\omega} (r) \equiv C^\pm_{\ell m\omega}\hat{R}^\pm_{\ell m\omega}(r)$. The constant matching coefficients are given by
\begin{align}
\label{MatchingCalc}
C_{\ell m\omega}^{\pm}& = \frac{1}{W_{\ell m\omega}} \frac{\hat{R}_{\ell m\omega}^{\mp}(r_\Omega) }{f_\Omega}\bar{G}_{\ell m\omega} \nonumber \\
& \qquad -\frac{1}{W_{\ell m\omega}} \frac{d}{dr}\left. \left(\frac{\hat{R}_{\ell m\omega}^{\mp}(r) }{f(r)}\right) \right\rvert_{r_\Omega} \bar{F}_{\ell m\omega} \nonumber \\
& \qquad + \frac{1}{W_{\ell m\omega}} \frac{d^2}{dr^2}\left. \left(\frac{\hat{R}_{\ell m\omega}^{\mp}(r) }{f(r)}\right) \right\rvert_{r_\Omega} \bar{H}_{\ell m\omega}.
\end{align}

In addition to the delta singularity, the resulting master function and its derivative have jump discontinuities at $r_\Omega$. The expressions for the jumps are obtained by substituting \eqref{eqn:formR} and \eqref{eqn:explicitsource} into \eqref{eqn:RWZfreq}. Defining $[[\psi(r_\Omega)]]_{\ell m \omega} \equiv R_{\ell m \omega} ^+(r_\Omega) -R_{\ell m \omega} ^-(r_\Omega)$ and $[[\psi'(r_\Omega)]]_{\ell m \omega} \equiv \partial_r R_{\ell m \omega} ^+(r_\Omega) - \partial_r R_{\ell m \omega} ^-(r_\Omega)$;
\begin{align}
f_\Omega^2 [[\psi(r_\Omega)]]_{\ell m \omega} &=\bar{F}_{\ell m \omega} + 3 \frac{f'_\Omega}{f_\Omega} \bar{H}_{\ell m \omega},\\
f_\Omega^2 [[\psi'(r_\Omega)]]_{\ell m \omega} &= \left( V_l(r_\Omega)-\omega^2 +2 (f'_\Omega)^2 - f_\Omega f''_\Omega \right)\frac{\bar{H}_{\ell m \omega}}{f_\Omega^2} \nonumber \\
 &\qquad +\bar{G}_{\ell m \omega}+ \frac{f'_\Omega}{f_\Omega}\bar{F}_{\ell m \omega},
\end{align}
where quantities with the subscript $\Omega$ are evaluated at $r_\Omega$, e.g. $f_\Omega\equiv f(r_\Omega)$.\\

Finally, obtaining the time domain master functions from their frequency domain counterparts is trivial for the case of circular equatorial orbits, since there is only a single frequency per $m$ mode and thus the time domain master functions are given by
\begin{equation}
\Psi_{\ell m}(t,r) = \psi_{\ell m\omega}(r) e^{-i \omega_m t}.
\end{equation}
In our two-timescale expansion, this instead becomes
\begin{equation}
\Psi_{\ell m}(t,r) = \psi_{\ell m\omega}(r) e^{-im \phi_p(t)}.
\end{equation}

\subsection{Metric reconstruction}
\label{sec:metric_reconstruction}

We next obtain the actual metric perturbation sourced by the stress-energy \eqref{T Taylor} due to a spinning point particle. This metric perturbation has two components: a \emph{radiative} piece which can be reconstructed from the RWZ master functions; and a \emph{completion} piece that fully captures the mass and angular momentum content of the perturbation. 

\subsubsection{Reconstruction from RWZ master functions}

The radiative part of the metric perturbation can be reconstructed by applying differential operators to the RWZ master functions along with source terms involving the stress-energy tensor. The metric perturbation one obtains is in the Regge-Wheeler/ Zerilli gauges. In terms of the complex vector $m^\mu = \frac{1}{\sqrt{2}r}\{0,0,1,i \csc \theta\}$ (with complex conjugate $\bar{m}^\mu$) the Regge-Wheeler gauge condition is equivalent to the conditions $h_{mm} = 0 = h_{t m} = h_{r m}$ in the even-parity sector and to the condition $h_{mm} = 0$ in the odd-parity sector. This reconstruction procedure is by now well-established and we omit the details here as they are given in full in Refs.~\cite{PhysRevD.82.084010} and \cite{pound2021black}.

\subsubsection{Metric completion at fixed frequency}
\label{sec:metric_completion}

The completion part of the metric perturbation can be obtained by solving the harmonic decomposed linearized Einstein equations directly for the $\ell=0$ and $\ell=1$ modes. The previous RWZ gauge conditions do not fully fix the gauge in this case since by definition $h_{mm} = 0$ for $\ell=0, 1$ and $h_{t m} = 0 = h_{r m}$ for $\ell=0$. Instead we fix the residual gauge freedom by working in a ``RWZ-like'' gauge (see Appendix \ref{sec:lorenzmono} for further details).

The metric completion pieces for a secondary with aligned spin in a circular orbit in a Schwarzschild background spacetime were first given in a RWZ-like gauge in \cite{Bini:2018zde}. These results were derived at fixed radius and their contribution to the redshift was later rewritten at fixed frequency. For completeness in Appendix \ref{sec:lorenzmono} we derive the completion pieces directly within a fixed frequency calculation.

The key result from this derivation is that for $\ell=0$ the only non-zero components of the metric perturbation for a spin-aligned secondary in a circular orbit in a Zerilli-like gauge are:
\begin{align}
h^{\ell=0}_{tt}&= \frac{2 \mu E }{r}\bigg\{\Theta[r-r_\Omega] \nonumber \\
 & \quad + \frac{rf}{r_\Omega f_\Omega}\left[1-\chi \mu \Omega \left(\frac{r_\Omega-3M}{r_\Omega-2M}\right) \right] \Theta[r_\Omega-r]\bigg\} \nonumber \\
 \label{eqn:monopole}
h^{\ell=0}_{rr}&= \frac{2 \mu E }{r f^2}\Theta[r-r_\Omega].
\end{align}\\
Note that the expression in Ref.~\cite{Bini:2018zde} for $h_{r r}$ featured a Dirac delta that does not appear when expressed at fixed frequency.

The $\ell=1$ contribution is comprised of the odd parity $m=0$ mode and the even parity $m=\pm 1$ modes. The even parity modes are pure gauge modes away from the worldine and also do not contribute to the redshift gauge invariant in Section \ref{sec:redshift}. The odd parity mode does contribute to the redshift invariant. In a Regge-Wheeler-like gauge, the only non-zero retarded metric perturbation component for a spin-aligned secondary in a circular orbit is
\begin{align}
\label{eqn:dipole}
h_{t\phi}&= -\frac{2 \mu J }{r}\bigg\{\Theta[r-r_\Omega] \nonumber \\
 &\quad + \frac{r^3}{r_\Omega^3} \left[1+\frac{3}{2} \chi \mu \frac{\Omega}{M}\left(r_\Omega-3M\right)\right]\Theta[r_\Omega-r] \bigg\}\sin^2(\theta).
\end{align}

\subsubsection{Evaluation on the worldline}

There is a subtlety in the fixed-frequency formulation in that when we wish to evaluate ``on the worldline'' (for example, to compute the local force or Detweiler's redshift) we must evaluate the metric perturbation at $r = r_p$. When evaluating a given expression, we must therefore substitute the expansion $r_p = r_\Omega + \e r_\chi$, re-expand and truncate in order to obtain a consistent result involving the metric perturbation and its derivatives evaluated at $r = r_\Omega$. An example of this final step is given in Sec.~\ref{sec:redshift}, in which which we use it when computing Detweiler's redshift.

\subsection{Gravitational wave fluxes}
\label{sec:flux}
The gravitational wave energy and angular momentum fluxes at $r_{*}=\pm \infty$ (more formally, at future null infinity and at the event horizon) are in general gauge invariant. They also involve only dissipative contributions to the metric perturbation and thus have the advantages of not requiring regularization and of involving a rapidly-convergent sum over spherical harmonic modes.\\

Provided the homogeneous radial solutions $\hat{R}_{\ell m\omega}^{\mp}(r)$ have been normalised as unit in-going/outgoing waves at $r_{*}=\mp \infty$, the specific (per unit $\mu$) energy and specific angular momentum fluxes (with respect to coordinate time $t$) are \cite{PhysRevD.82.084010} 
\begin{equation*}
\dot{E}^{\pm}=\displaystyle\sum_{\ell=2}^{\infty} \displaystyle \sum^{\ell}_{m=-\ell} \dot{E}_{\ell m}^{\pm}, \quad \dot{J}^{\pm}=\displaystyle\sum_{\ell=2}^{\infty} \displaystyle \sum^{\ell}_{m=-\ell} \dot{J}_{\ell m}^{\pm }.
\end{equation*}
where
\begin{align}
\label{eqn:Eflux}
\dot{E}_{\ell m\omega}^{\pm }=\frac{1}{64 \pi \mu} \frac{(\ell+2) !}{(\ell-2) !} \omega_{m}^{2}\left|C_{\ell m\omega}^{\pm}\right|^{2}, \\
\label{eqn:Jflux}
\dot{J}_{\ell m\omega}^{\pm }=\frac{m}{64 \pi \mu} \frac{(\ell+2) !}{(\ell-2) !} \omega_{m}\left|C_{\ell m\omega}^{\pm}\right|^{2},
\end{align}
are the harmonic modes of the energy and angular momentum fluxes at $r_{*}=\pm \infty$ respectively. The $C_{\ell m\omega}^{\pm}$ are those given by Eqn.~\eqref{MatchingCalc} and also depend on $\mu$. Note that neither the zero frequency $m=0$ modes nor the conservative $\ell=0$ and $\ell=1$ modes contribute to the fluxes so the mode-sum starts at $\ell=2$.

\section{Detweiler's redshift invariant}
\label{sec:redshift}

Detweiler's gauge invariant redshift --- $z\equiv\frac{d\hat\tau}{dt}=1/\hat u^t$--- was generalised to the case of an (anti-)aligned spinning secondary in a circular orbit on a Schwarzschild background spacetime in Ref.~\cite{Bini:2018zde}.  Written in a ``mixed" form in terms of the invariant frequency variable $y=M/r_\Omega$ and orbital radius $r_p=r_\Omega+\e r_\chi(y)$, the explicit form of the redshift is given by\footnote{This expression is valid for (anti-)aligned spinning secondaries in a circular orbit parameterised at fixed frequency only.}
\begin{equation}
\label{eqn:spinredshift}
z(y,r_p,\e)= z_{(0)}(y) + \e z_{(1)}(y,r_p) +\e^2 z_{(2)}(y,r_p) +\mathcal{O}(\e^3),
\end{equation}
where we have defined
\begin{align}
\label{eq:redshiftcomponentsA}
z_{(0)}(y) &\equiv \sqrt{1-3y},\\
z_{(1)}(y,r_p) &\equiv -\frac{1}{2\sqrt{1-3y}}h^{\rm R \, (1)}_{kk}(r_p),\label{z1}
\end{align}
and as usual we restrict our analysis of the $O(\e^2)$ term to spin effects only, which contribute
\begin{multline}
z_{(2)}(y,r_p) \equiv -\frac{1}{2\sqrt{1-3y}}\Big[ h^{\rm R \, (\chi)}_{kk}(r_p) \\
+ \mu \chi y^{1/2}\partial_r h^{\rm R \, (1)}_{kk}(r_p)\Big].
\label{eq:redshiftcomponentsB}
\end{multline}
We have also defined the helical Killing vector $k^\alpha\equiv \xi_{(t)}^{\alpha}+\Omega \xi_{(\phi)}^{\alpha}$ and $h^{\rm R}_{kk}\equiv h^{\rm R}_{\alpha \beta} k^{\alpha}k^{\beta}$. The radial derivative term in Eq.~\eqref{eq:redshiftcomponentsB} follows from the MPD equations in the perturbed spacetime, and can be interpreted as ensuring the redshift is gauge invariant through linear order in spin.

From Eq.~\eqref{eqn:spinredshift} we can obtain a simpler expression in which $z$ is fully expanded in powers of $\e$ at fixed $\Omega$:
\begin{equation}
\label{eqn:spinredshift v2}
z(y,\e) = z_{(0)}(y) + \e z_{(1)}(y) +\e^2 z_{(2)}(y)+\mathcal{O}(\e^3).
\end{equation}
The single-argument functions here are related to the two-argument functions in Eq.~\eqref{eqn:spinredshift} by $z_{(1)}(y) = z_{(1)}(y,r_\Omega)$ and $z_{(2)}(y) = z_{(2)}(y,r_\Omega) +r_\chi(y) \partial_{r_\Omega} z_{(1)}(y,r_\Omega)$.
To understand the expansion in more detail, note that in the equations above, $h^{\rm R \, (1)}_{kk}(r_p)$ and $h^{\rm R \, (\chi)}_{kk}(r_p)$ are evaluated at the field point $r=r_p$, but they also denote fields that are generated by a particle at $r_p$. 
We can make this explicit by writing the metric perturbation's dependence on the field point $x^{\alpha}$ and on the secondary body's worldline $z^{\alpha'}$ as $h_{kk}=h_{kk}(z^{\alpha'}, x^{\alpha})$. 
Substituting $r_p=r_\Omega+\e r_\chi$ and expanding both arguments, we obtain 
\begin{align}\label{eqn:hR expansion}
h^{\rm R \, (1)}_{kk}(z^{\alpha'}, z^{\alpha}) &= h^{\rm R \, (1)}_{kk}(z^{\alpha'}_0, z^{\alpha}_0) +\e r_\chi \partial_{r_\Omega'} h_{kk}^{\mathrm{R}\,(1)}(z^{\alpha'}_0, z^{\alpha}_0) \nonumber\\ 
&\quad+ \e r_\chi\partial_{r_\Omega} h_{kk}^{\mathrm{R}\,(1)}(z^{\alpha'}_0, z^{\alpha}_0) +\mathcal{O}(\e^2),
\end{align}
and $h^{\rm R \, (\chi)}_{kk}(z^{\alpha'}, z^{\alpha}) = h^{\rm R \, (\chi)}_{kk}(z_0^{\alpha'}, z^{\alpha}_0) + \mathcal{O}(\e)$. Inspecting Eqs.~\eqref{eqn:spinredshift}, \eqref{z1}, and \eqref{eq:redshiftcomponentsB} again, and noting $r_\chi=- \mu \chi y^{1/2}$, we see that the final term in Eq.~\eqref{eqn:hR expansion} cancels the term involving $\mu \chi y^{1/2}\partial_r h^{\rm R \, (1)}_{kk}$ in $z_{(2)}(y,r_p)$. The second term in Eq.~\eqref{eqn:hR expansion} combines with the $h^{\rm R \, (\chi)}_{kk}$ term in  $z_{(2)}(y,r_p)$ to give the total linear-in-spin contribution to $h^{\rm R(2)}_{kk}$: 
\begin{multline}
h_{kk}^{\mathrm{R}(2, \chi)}(z_0^{\alpha'}, z_0^{\alpha}) = h_{kk}^{\mathrm{R}(\chi)}(z^{\alpha'}_0, z^{\alpha}_0) \\+r_\chi \partial_{r_\Omega'} h_{kk}^{\mathrm{R}(1)}(z^{\alpha'}_0, z^{\alpha}_0).\label{eq:hspinpart}
 \end{multline}
The two terms in $h_{kk}^{\mathrm{R}(2, \chi)}$ correspond to the two terms in $T^{\mu \nu}_{(2)}$ from Eq.~\eqref{T2 Taylor}.  

The explicit version of Eq.~\eqref{eqn:spinredshift v2} is then simply
\begin{equation}
\label{eqn:spinredshift v3}
z=\sqrt{1-3y}-\frac{1}{2\sqrt{1-3y}}\left[\e h^{\rm R\,(1)}_{kk}(r_\Omega)+ \e^2 h_{kk}^{\rm R \,(2,\chi)}(r_\Omega)\right],
\end{equation}
where we only keep the linear-in-spin second-order terms,
\begin{equation}
\label{eqn:spinredshift 2 chi}
z_{(2,\chi)}=-\frac{h_{kk}^{\rm R \,(2,\chi)}(r_\Omega)}{2\sqrt{1-3y}}.
\end{equation}
This formula can also be deduced from Eq.~(100) in Ref.~\cite{Pound:2014koa}, which is valid for any radial perturbing force and includes the complete order-$\e^2$ term.

\section{Regularization}
\label{sec:reg}

A consequence of modelling a compact object by a Dirac delta function and its derivatives is that the retarded metric perturbation is singular on the worldline. This leads to discontinuities in the RWZ master functions across the worldline and the sum over spherical harmonic modes will not converge there. In fact, equation~\eqref{eqn:formR} shows that for a spinning body modelled in this way there is also a delta singularity in the master functions on the worldline \footnote{As $X_{\ell m\omega}\equiv \frac{\bar{H}_{\ell m\omega}}{f^2_{\Omega}}$ and $\bar{H}^{\rm even}_{\ell m\omega}=0$ when parameterised at fixed frequency, the delta singularity on the worldline vanishes in the even parity master function.}. These spurious divergences are not fundamental, and can be unambiguously avoided by a more careful treatment that uses matched asymptotic expansions instead of distributions \cite{pound2021black}. The net result of such an analysis is that at leading order in perturbation theory (including subleading order spin terms) we recover the point particle approximation with distributional sources, along with a well-defined regularization procedure that involves subtracting an appropriate singular field from the retarded field to produce a so-called residual field. That is, 
\begin{equation}
\label{eq:hR}
h_{\mu \nu}^{\rm R} = h_{\mu \nu}^{\rm ret} - h_{\mu \nu}^{\rm S},
\end{equation}
where the superscripts `$\rm R$' and `$\rm S$' denote the regular and singular pieces, respectively. It is this residual field that appears in local quantities evaluated on the worldline.

The spin's contribution to the local field $h_{\mu \nu}^{(2)\rm ret}$ near the particle was derived in Ref.~\cite{Pound:2009sm} as a local expansion in powers of distance from the worldline, through order (distance)$^0$, and in Ref.~\cite{Pound:2012dk} through linear order in distance. Ref.~\cite{Pound:2012dk} also defined a split into singular and regular pieces. However, the field was expressed in a local coordinate system; some additional work is required to put  $h_{\mu \nu}^{\rm S(2,\chi)}$ in a practical form that can be used to calculate $h^{\rm R}_{kk}$. Moreover, as explained in Sec.~\ref{sec:eom}, it was not shown that the regular field defined in Ref.~\cite{Pound:2012dk} is the one that enters into the equations of motion.

Here, we instead adopt the Detweiler-Whiting \cite{Detweiler2002} approach to definining a singular-regular split. While it has not been rigorously shown that the resulting regular metric produces the correct force for a spinning body, we again point out that Harte's \cite{Harte:2011ku} choice of effective metric \textit{has} been shown to do so and is consistent with Detewiler and Whiting's choice through linear order in the secondary's mass and spin --- to which our calculations are restricted. The resulting approximated singular field is also consistent with that of Ref.~\cite{Pound:2012dk} \textit{at least} through the orders required for the calculation of Detweiler's redshift.

\subsection{Detweiler-Whiting singular field}
\label{sec:DW_sing_field}

In defining a singular-regular split our main criteria are that the singular field has the same singular structure as the retarded field in the vicinity of the secondary's worldline, and that it must not contribute to the equations of motion.  Detweiler and Whiting \cite{Detweiler2002} have shown that in the non-spinning case an appropriate singular field can be defined in terms of a Green function decomposition, which is best understood in the Lorenz gauge. Furthermore, the singular field they identified has the property that when subtracted from the retarded field, the residual regular field satisfies the homogeneous Lorenz-gauge wave equation.

The \textit{trace-reversed} Detweiler-Whiting singular field is defined by
\begin{equation}
\label{lorenzpert}
\bar{h}_{\alpha \beta}^{\rm S}(x)=4 \int G^{\rm S}_{\alpha\beta\alpha'\beta'}(x,x')T^{\alpha'\beta'}(x')\sqrt{-g'}d^4x',
\end{equation}
where $G^{\rm S}_{\alpha\beta\alpha'\beta'}(x,x')$ is the Detweiler-Whiting singular Green function.
Within a normal neighbourhood the singular Green function can be expressed in its Hadamard form \cite{Poisson2011,Detweiler2002},
\begin{equation}
\label{hadamard}
G_{\alpha\beta\alpha'\beta'}^{\rm S}= \frac{1}{2}\Big[U_{\alpha\beta\alpha'\beta'} \delta\left(\sigma\right)
 +V_{\alpha\beta\alpha'\beta'} \theta\left(\sigma\right)\Big],
\end{equation}
where $\sigma(x, x^{\prime})$ is the Synge world function, and where $U(x, x^{\prime})_{a b a^{\prime} b^{\prime}}$ and $V(x, x^{\prime})_{a b a^{\prime} b^{\prime}}$ are symmetric bi-tensors.
The singular metric perturbation for a spinning body in the Lorenz gauge can be expressed in terms of these fundamental bi-tensors by substituting the stress-energy in Eqs.~\eqref{split}, \eqref{stressenergy} and \eqref{stressenergyspin}, and the singular Green function in Eq.~\eqref{hadamard} into Eq.~\eqref{lorenzpert}. Considering the mass-monopole and spin-dipole contributions to the singular field separately,
\begin{equation*}
\bar{h}_{\alpha\beta}^{\rm S} = \bar{h}_{\alpha\beta}^{\rm S (\mu)} + \bar{h}_{\alpha\beta}^{\rm S (\chi)},
\end{equation*}
where
\begin{align*}
\bar{h}_{\alpha\beta}^{\rm S (\mu / \chi)}(x)=4 \int G^{\rm S}_{\alpha\beta\alpha'\beta'}(x,x')T^{(\mu/ \chi)\alpha'\beta'}(x')\sqrt{-g'}d^4x',
\end{align*}
the mass-monopole contribution to the singular field is well known \cite{Detweiler2002, PhysRevD.86.104023} and is given by
\begin{align}
\label{eqn:monopole_sing}
\bar{h}_{\alpha\beta}^{\rm S (\mu)}(x) &=2 \mu \left. \left[\frac{U\left(x, x'\right)_{\alpha\beta\alpha'\beta'}u^{\alpha'} u^{\beta'}}{\left|\sigma_{\gamma'}u^{\gamma'}\right|}\right] \right|_{x'=x_{A/R}} \nonumber \\
& \qquad
 +  2 \mu \int_{\tau_{R}}^{\tau_{A}} V\left(x, z(\tau') \right)_{\alpha\beta\alpha'\beta'}u^{\alpha'}u^{\beta'} d \tau'.
\end{align}
Here, the shorthand $[\cdots]|_{x'=x_{A/R}} $ corresponds to $ [\cdots]|_{x'=x_{A}} + [\cdots]|_{x'= x_{R}}$ where $x_A$ is the advanced point at which the future lightcone of $x$ intersects the worldline and $x_R$ is the retarded point at which the past lightcone of $x$ intersects the worldline. The spin-dipole contribution (derived in Appendix \ref{sec:singular-dipole}) is given by
\begin{align}
\bar{h}_{\alpha\beta}^{\rm S (\chi)}(x) &= 2 \mu^2 \Bigg[ \bigg( \frac{u^{\rho'}\nabla_{\rho'}U_{\alpha\beta\alpha'\beta'}  \sigma_{\rho'}+U_{\alpha\beta\alpha'\beta'}  u^{\kappa'}\sigma_{\rho' \kappa'}}{\sigma_{\delta'}u^{\delta'}} \nonumber \\
 &  - ( \nabla_{\rho'}  U_{\alpha\beta\alpha'\beta'} + V_{\alpha\beta\alpha'\beta'} \sigma_{\rho'}) \nonumber \\
 &  -  \frac{U_{\alpha\beta\alpha'\beta'} \sigma_{\rho'}\sigma_{\gamma' \kappa'}u^{\gamma'}u^{\kappa'}}{(\sigma_{\delta'}u^{\delta'})^2}\bigg)  \frac{u^{(\alpha'}\tilde{S}^{\beta') \rho'}}{|\sigma_{\gamma'}u^{\gamma'}|} \Bigg] \Bigg|_{x'=x_{A/R}}\nonumber \\
\label{eqn:dipole_sing}
&  -  2 \mu^2  \int_{\tau_{R}}^{\tau_{A}} \nabla_{\rho'}  V\left(x, z(\tau')\right)_{\alpha\beta\alpha'\beta'} u^{(\alpha'}\tilde{S}^{\beta') \rho'} d \tau',
\end{align}
where this result is not yet specialised to a given spacetime, spin alignment or orbital configuration.

\subsection{Tensor harmonic regularisation parameters}
\label{sec:TensorRPs}

The forms of equation~\eqref{eqn:monopole_sing} and \eqref{eqn:dipole_sing} are not yet suitable to subtract the singular metric perturbation from the retarded metric perturbation on the worldline, since subtracting infinity from infinity is not well defined. Instead, we use the mode-sum regularisation approach originally pioneered in the case of a scalar charged particle in Schwarzschild spacetime in \cite{Barack1999,Barack2001}. The idea behind the approach is that while the mode-sum producing the retarded metric perturbation is singular on the secondary body's worldline, the discrete modes in the sum are themselves finite. Thus, one can subtract off the singular metric perturbation mode-by-mode to leave a regular metric perturbation for which the sum converges to a finite result.

In fact, in the mode sum approach only an approximation to the singular metric perturbation is required. The mode decomposed singular field can be represented as an infinite series of so-called regularisation parameters and only the first few regularisation parameters need to be included to achieve a finite mode-sum; the inclusion of successive regularisation parameters (associated with higher-order approximations to the singular field) only serves to speed up the rate of convergence of the sum. In the case of the redshift invariant for a spinning body discussed in Sec.~\ref{sec:redshift}, the subtraction of the first two regularisation parameters is sufficient to render the mode-sum convergent.

In order to derive mode-sum regularisation parameters, we start from a suitable local expansion of the singular field. High order expansions of the singular field for a non-spinning secondary in Schwarzschild spacetime were produced in \cite{PhysRevD.86.104023} and used to derive regularisation parameters for various self-force quantities. That work was later extended to Kerr spacetime \cite{Heffernan2012} and then to accelerated bodies in the scalar-field case \cite{Heffernan2017}. Applying the covariant expansion techniques developed in those earlier works to the singular field in Eqs.~\eqref{eqn:monopole_sing} and \eqref{eqn:dipole_sing} and imposing the condition $u_{\alpha}\tilde{S}^{\alpha \beta}=\mathcal{O}(\e^2)$ leads to an approximation for the Detweiler-Whiting singular field given by
\begin{align}
\label{eqn:sing_field_approx}
\bar{h}_{\mu \nu}^{\rm S}&=\mu^2 \frac{4g_{\mu}^{\bar{\alpha}}g_{\nu}^{\bar{\beta}}u_{(\bar{\alpha}}\tilde{S}_{\bar{\beta})\bar{\gamma}}\sigma^{\bar{\gamma}}}{\bar{\lambda}^2 \bar{s}^3} + \mu \frac{4g_{\mu}^{\bar{\alpha}}g_{\nu}^{\bar{\beta}}u_{\bar{\alpha}}u_{\bar{\beta}}}{\bar{\lambda} \bar{s}} \nonumber \\
& \qquad +\bar{h}_{\mu \nu}^{\rm S(0)}+\bar{\lambda}\bar{h}_{\mu \nu}^{\rm S(1)}+\mathcal{O}(\bar{\lambda}^2,\e^2),
\end{align}
where $\bar{s}\equiv (g_{\bar{a} \bar{b}}+u_{\bar{a}}u_{\bar{b}})\sigma^{\bar{a}}\sigma^{\bar{b}}$.
Here, $\bar{\lambda}$ is simply used as an order counting parameter measuring distance from the worldline. The bar over the indices represent evaluation at $\bar{x}$, an arbitrary point on the worldline. The leading-order spin term, of order $\bar{\lambda}^{-2}$, has been independently derived using matched asymptotic expansions \cite{pound2021black}, and the first subleading term, of order $\bar{\lambda}^{-1}$, derives purely from the mass-monopole and is already well-established \cite{PhysRevD.86.104023}. The subleading terms, of order $\bar{\lambda}^0$ and $\bar{\lambda}^1$, are expressed in covariant form here (see Appendix~\ref{sec:SingulFieldExpansion}) for the first time, though their equivalents were produced in Ref.~\cite{Pound:2012dk} in Fermi-Walker coordinates.\\

The next step in deriving regularization parameters is to perform a coordinate series expansion of Eq.~\eqref{eqn:sing_field_approx} and to decompose the result into a basis of spherical-harmonic modes.
As our retarded metric perturbation is decomposed into a basis of scalar, vector and tensor spherical harmonics, to use the traditional mode-sum approach of using scalar-harmonic regularisation parameters would require us to project our tensor harmonic modes onto a basis of scalar spherical harmonics. To avoid this issue, Wardell and Warburton \cite{PhysRevD.92.084019} derived a tensor-harmonic mode-sum regularisation procedure and applied it to the non-spinning case. We now follow their methodology to derive tensor-harmonic regularisation parameters for the case of a spinning body.  The process is technically involved, but follows exactly the procedure described in Ref.~\cite{PhysRevD.92.084019} so we only briefly summarise the key results here.

The final form of the tensor-harmonic-mode decomposed singular field is
\begin{equation}
\label{eqn:rp_notation}
h_{\mu \nu }^{\rm S,\ell}= \pm(2\ell+1)h_{\mu \nu}^{[-1]}+h_{\mu \nu}^{[0]}+\mathcal{O}(\ell^{-2}),
\end{equation}
where $h_{\mu \nu}^{[-1]}$ and $h_{\mu \nu}^{[0]}$ are the leading two regularisation parameters. 
We then obtain the regular metric perturbation via a mode-sum regularization procedure,
\begin{equation}
\label{eqn:regmetric}
h_{\mu \nu}^{\rm R}=\sum_{\ell=0}^{\infty}\left[h_{\mu \nu}^{{\rm ret},\ell} \mp (2\ell+1)h_{\mu \nu}^{[-1]}-h_{\mu \nu}^{[0]}\right].
\end{equation}

At this point, it is important to point out that since Eqs.~\eqref{eqn:monopole_sing} and \ref{eqn:dipole_sing} are derived in the Lorenz gauge, the regularisation parameters for the components of the metric perturbation are, in general, only suitable for self-force calculations in the Lorenz gauge. It is possible to transform the regularisation parameters to the Regge-Wheeler gauge \cite{PhysRevD.99.124046}, but in our case this is not necessary as we are ultimately interested in computing gauge invariant quantities and thus may use the parameters calculated in the Lorenz gauge. 
In particular, inserting the regularization parameters for the Lorenz gauge metric perturbation into the expression for the redshift, Eq.~\eqref{eqn:spinredshift v3}, we arrive at a mode-sum formula for the gauge-invariant redshift,
\begin{equation}
\label{eqn:regz}
z=\sum_{\ell=0}^{\infty}\left[z^{{\rm ret},\ell} \mp (2\ell+1)z^{[-1]}-z^{[0]}\right].
\end{equation}
with regularization parameters given by
\begin{subequations}
\begin{align}
 z^{[-1]} &= 
 \begin{dcases}
 \mu^2 \chi \frac{M^{1/2}(r_\Omega-3M)}{2r_\Omega^{5/2} (r_\Omega -2M)}, & \ell\geq2,\\
  \mu^2 \chi \frac{M^{1/2}}{2 r_\Omega^{5/2}}, & \ell=1,\\
 - \mu^2 \chi \frac{M^{1/2}}{2 r_\Omega^{5/2}}, & \ell=0,
 \end{dcases} \\
z^{[0]}&=- \frac{ 2 \mu \mathcal{K} (r_\Omega^2 f_\Omega - 16 M r_\Omega f_\Omega \Lambda_1 + 16 M^2 \Lambda_2)}{\pi r_\Omega^{3/2} (r_\Omega - 3 M)(r_\Omega - 2 M)^{1/2} } \nonumber \\
&+ \frac{\mu^2 \chi }{\pi M^{1/2} r_{\Omega }^3 \left(r_{\Omega }-3M\right)(r_\Omega - 2 M)^{1/2}} \times \nonumber \\
&\quad\Big\{ M \big[2 \mathcal{K }\left(2 r_{\Omega }-3M\right) \left(4M-r_{\Omega }\right) \nonumber \\
& \qquad\quad +\mathcal{E} \left(9 M-5 r_{\Omega }\right) \left(2 M-r_{\Omega }\right)\big]\nonumber \\
&\qquad+ 8\Lambda_1 \big[\mathcal{E} \left(2 M-r_{\Omega }\right) \left(13 M^2-9 M r_{\Omega }+2 r_{\Omega }^2\right) \nonumber \\
& \qquad\quad +2 \mathcal{K} \left(-14 M^3+16 M^2 r_{\Omega }-7 M r_{\Omega }^2+r_{\Omega }^3\right)\big] \nonumber\\
&\qquad+ 16 \Lambda_2 M \big[\mathcal{E} \left(11 M^2-9 M r_{\Omega }+2 r_{\Omega }^2\right) \nonumber\\
& \qquad\quad-2\mathcal{K} \left(5 M^2-5 M r_{\Omega }+r_{\Omega }^2\right)\big]\Big\}.
\end{align}
\end{subequations}
Here, $\mathcal{K} \equiv \int_{0}^{\pi / 2}\left(1-\frac{M}{r_{\Omega}-2 M} \sin ^{2} x\right)^{-1 / 2} d x$ and $\mathcal{E} \equiv \int_{0}^{\pi / 2}\left(1-\frac{M}{r_{\Omega}-2 M} \sin ^{2} x\right)^{1 / 2} d x$ are elliptic integrals of the first and second kind, respectively, and we have introduced $\Lambda_{1} \equiv \frac{\ell(\ell+1)}{(2 \ell-1)(2 \ell+3)}$ and $\Lambda_{2} \equiv \frac{(\ell-1) \ell(\ell+1)(\ell+2)}{(2 \ell-3)(2 \ell-1)(2 \ell+3)(2 \ell+5)}$.

\section{Numerical results}
\label{sec:results}

Our numerical results naturally divide into three sections. In Section \ref{sec:fluxresults} we demonstrate the flux balance law, showing agreement between the asymptotic gravitational wave energy fluxes and the local rate of change of energy. In Section \ref{sec:redshiftresults} we give results for the Detweiler's redshift invariant. In Section \ref{sec:waveforms} we produce gravitational waveforms that are complete at adiabatic order and include post adiabatic spin effects. Unless otherwise stated, all numerical results are given adimensionalized in $M$, $\mu$ and $\chi$.

\subsection{Flux balance}
\label{sec:fluxresults}
In Table~\ref{tab:flux}, we reproduce the flux balance law results of \cite{Akcay:2019bvk} at fixed frequency, demonstrating Eqn.~\ref{balance law} by comparing the asymptotic energy flux at $r_*=\pm \infty$ with the rate of change of local energy at the worldline. In all cases we have summed up to $\ell_{\mathrm{max}}=30$ and have set the tolerances in our numerical integration of the retarded field equations such that the local rate of change of energy is accurate in all digits shown. 
We find that the flux balance law is satisfied to all significant digits in our calculation, with the asymptotic fluxes agreeing with the local rate of change of energy to an absolute accuracy of $10^{-28}$ or better.
\begin{table*}[htb]
\centering
$\begin{array}{c|ccc|c|c}
y& \dot{E}_0&\dot{E}^-_\chi&\dot{E}^+_\chi&d E/d\tau _\chi& \Delta^{\mathrm{rel}}_\chi\\
\hline
0.2 & 2.79273701868\times 10^{-3} & 3.77193403195\times 10^{-7} & -6.10406021099\times 10^{-4} & -9.64540266941\times 10^{-4} & 1.210\times
   10^{-28} \\
 0.18 & 1.46844806236\times 10^{-3} & 7.60541476292\times 10^{-8} & -2.60585846715\times 10^{-4} & -3.84100734136\times 10^{-4} & 1.029\times
   10^{-28} \\
 0.16 & 7.46754277822\times 10^{-4} & 1.08980506901\times 10^{-8} & -1.05064301974\times 10^{-4} & -1.45682859427\times 10^{-4} & 2.6\times
   10^{-30} \\
 0.14 & 3.58765894169\times 10^{-4} & 8.06926440798\times 10^{-10} & -3.89407471318\times 10^{-5} & -5.11306464414\times 10^{-5} & 1.7713\times
   10^{-28} \\
 0.12 & 1.58228153292\times 10^{-4} & -6.53905197867\times 10^{-11} & -1.28067951208\times 10^{-5} & -1.60085756392\times 10^{-5} & 1.144\times
   10^{-29} \\
 0.1 & 6.15163167846\times 10^{-5} & -2.66993570598\times 10^{-11} & -3.54917559346\times 10^{-6} & -4.24210812069\times 10^{-6} & 4.627\times
   10^{-29} \\
 0.09 & 3.59063362311\times 10^{-5} & -1.01476993750\times 10^{-11} & -1.71031987628\times 10^{-6} & -2.00178988091\times 10^{-6} & 4.20\times
   10^{-30} \\
 0.08 & 1.97579085327\times 10^{-5} & -3.10096178158\times 10^{-12} & -7.62066085171\times 10^{-7} & -8.74153307983\times 10^{-7} & 4.4989\times
   10^{-29} \\
 0.07 & 1.00797672995\times 10^{-5} & -7.55072229184\times 10^{-13} & -3.07211805328\times 10^{-7} & -3.45641134719\times 10^{-7} & 1.45815\times
   10^{-28} \\
 0.06 & 4.65287054407\times 10^{-6} & -1.40588119661\times 10^{-13} & -1.08551794346\times 10^{-7} & -1.19875558331\times 10^{-7} & 1.11333\times
   10^{-28} \\
 0.05 & 1.87147091142\times 10^{-6} & -1.85060798132\times 10^{-14} & -3.20089991680\times 10^{-8} & -3.47186542918\times 10^{-8} & 4.890\times
   10^{-30} \\
 0.04 & 6.15791960326\times 10^{-7} & -1.49663127140\times 10^{-15} & -7.25545365705\times 10^{-9} & -7.73434118125\times 10^{-9} & 2.074\times
   10^{-29} \\
 0.03 & 1.47265886605\times 10^{-7} & -5.67900033298\times 10^{-17} & -1.08380957000\times 10^{-9} & -1.13614119765\times 10^{-9} & 4.583\times
   10^{-30} \\
 0.02 & 1.96245785614\times 10^{-8} & -5.49135672040\times 10^{-19} & -7.55124235208\times 10^{-11} & -7.78851185422\times 10^{-11} &
   3.2001\times 10^{-29} \\
 0.015 & 4.69335489271\times 10^{-9} & -2.02390121354\times 10^{-20} & -1.14903370686\times 10^{-11} & -1.17579357813\times 10^{-11} &
   1.0545\times 10^{-28} \\
 0.01 & 6.23820347340\times 10^{-10} & -1.91947958972\times 10^{-22} & -8.14067891602\times 10^{-13} & -8.26560712092\times 10^{-13} &
   9.494\times 10^{-30} \\
\end{array}$
\caption{\label{tab:flux} Total non-spin energy flux ($\dot{E}_0$), linear-in-spin contributions to the asymptotic energy flux through the event horizon ($\dot{E}^-_\chi$) and future null infinity ($\dot{E}^+_\chi$), local rate of change of energy, $d E/d\tau_\chi$, and relative error in the linear-in-spin flux balance, $\Delta^{\mathrm{rel}}_\chi \equiv \left|1-\frac{u^t \dot{E}_\chi}{d E/d\tau_\chi} \right|$, as a function of the frequency of the circular orbit (represented by $y$).}
\end{table*}

\subsection{Detweiler's redshift invariant}
\label{sec:redshiftresults}
\begin{figure}[htb]
\includegraphics[width=0.95\columnwidth]{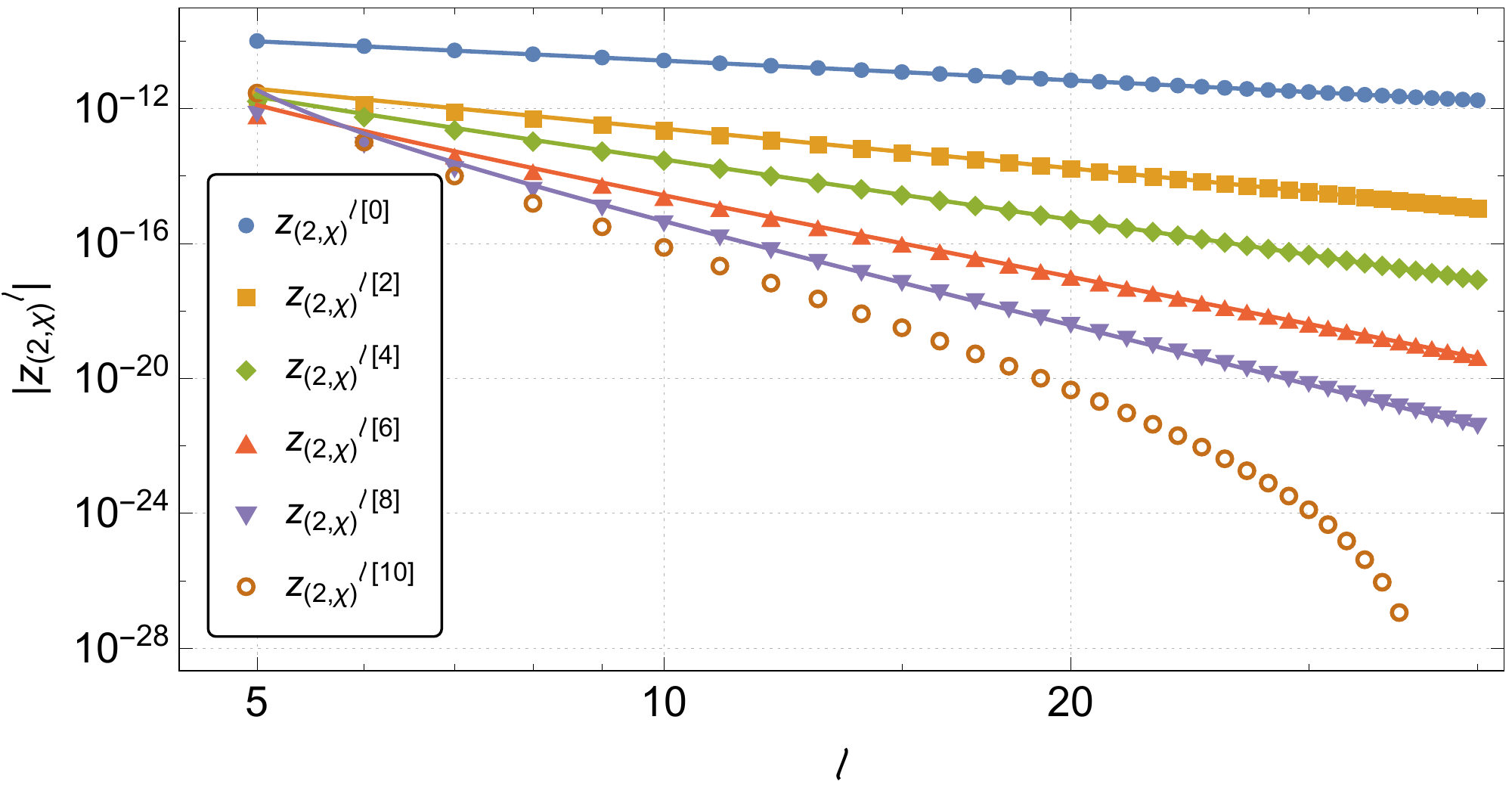}
\caption{An example of the mode-sum regularisation procedure used to calculate the linear-in-spin part of the redshift, $z_{(2, \chi)}$, for $r_\Omega=100M$. The data points show the absolute value of the successively regularized $\ell$ modes of $z_{(2, \chi)}$ when using different numbers of regularization parameters; the joined lines are the fitted regularisation parameters. The leading behaviour in $\ell$ of the regularized $z_{(2, \chi)}$ modes goes as the next leading regularisation parameter that was not included in the regularisation scheme. Hence the joined lines overlay the plotted points and qualitatively verify the fitted parameters.}
\label{fig:regparamsfit}
\end{figure}

In Table~\ref{tab:redshift}, we give results for Detweiler's redshift, which is a conservative gauge invariant quantity and requires regularization. While in Sec.~\ref{sec:TensorRPs} we have only derived the first two regularisation parameters analytically, we can improve the convergence of the mode-sum by fitting for higher regularisation parameters (see Fig.~\ref{fig:regparamsfit}) whose successive $\ell$ dependence is well known \cite{PhysRevD.92.084019}. We can confidently subtract the fitted parameters since we know their contribution formally sums to zero in the complete mode-sum.
\begin{table*}[htb]
\centering
$\begin{array}{c|cc|cc}
r_\Omega & z_{(1)} & z_{(2, \chi)} & \Delta_{z}^{(1)}& \Delta_{z}^{(2, \chi)} \\
\hline
 4 & 3.0467428778824171300\times 10^{-1} & -1.4586300570011681310\times 10^{-2} & -6.421372315\times 10^{-11} & 1.040868132\times 10^{-10} \\
 5 & 1.8666094967945293038\times 10^{-1} & -1.4359997170303332565\times 10^{-3} & -8.501492777\times 10^{-13} & 6.285718934\times 10^{-13} \\
 6 & 1.4801375464498224547\times 10^{-1} & 1.2778851458222201614\times 10^{-4} & -8.494359409\times 10^{-14} & 4.205017886\times 10^{-14} \\
 7 & 1.2619858710413698659\times 10^{-1} & 3.2895525026652979198\times 10^{-4} & -1.486953002\times 10^{-14} & 6.440450797\times 10^{-15} \\
 8 & 1.1107483972099400145\times 10^{-1} & 3.0130832079882955894\times 10^{-4} & -2.789239601\times 10^{-15} & 1.538046174\times 10^{-15} \\
 9 & 9.9573739278604293268\times 10^{-2} & 2.4150088443239778085\times 10^{-4} & -7.542900971\times 10^{-17} & 4.914937959\times 10^{-16} \\
 10 & 9.0385592074434347608\times 10^{-2} & 1.8779394202853647999\times 10^{-4} & 5.560967658\times 10^{-16} & 1.940250420\times 10^{-16} \\
 11 & 8.2817927552432716594\times 10^{-2} & 1.4571779302612887645\times 10^{-4} & 6.457897963\times 10^{-16} & 8.995489317\times 10^{-17} \\
 12 & 7.6451679289700237541\times 10^{-2} & 1.1390103980806758238\times 10^{-4} & 5.909892489\times 10^{-16} & 4.722720931\times 10^{-17} \\
 13 & 7.1010239380097475857\times 10^{-2} & 8.9977582894836949305\times 10^{-5} & 5.057588184\times 10^{-16} & 2.730507042\times 10^{-17} \\
 14 & 6.6300106250377251019\times 10^{-2} & 7.1894261756329048211\times 10^{-5} & 4.235268423\times 10^{-16} & 1.701375794\times 10^{-17} \\
 15 & 6.2180255400520542626\times 10^{-2} & 5.8095022692680346999\times 10^{-5} & 3.528715259\times 10^{-16} & 1.123780963\times 10^{-17} \\
 16 & 5.8544734161342785070\times 10^{-2} & 4.7448661525134242815\times 10^{-5} & 2.945648894\times 10^{-16} & 7.770696488\times 10^{-18} \\
 17 & 5.5312035461636850090\times 10^{-2} & 3.9141832229418841873\times 10^{-5} & 2.471167798\times 10^{-16} & 5.572694481\times 10^{-18} \\
 18 & 5.2418249523532483206\times 10^{-2} & 3.2588897056415252246\times 10^{-5} & 2.086117042\times 10^{-16} & 4.115745383\times 10^{-18} \\
 19 & 4.9812462417061066657\times 10^{-2} & 2.7365239308320767569\times 10^{-5} & 1.772872625\times 10^{-16} & 3.113963219\times 10^{-18} \\
 20 & 4.7453560812119756566\times 10^{-2} & 2.3160083573548132458\times 10^{-5} & 1.516776702\times 10^{-16} & 2.403886377\times 10^{-18} \\
 25 & 3.8367861668465180133\times 10^{-2} & 1.1094904818711662769\times 10^{-5} & 7.613023266\times 10^{-17} & 8.160586074\times 10^{-19} \\
 30 & 3.2200482214638454294\times 10^{-2} & 6.0251503215045887450\times 10^{-6} & 4.312790601\times 10^{-17} & 3.486595670\times 10^{-19} \\
 35 & 2.7740018849300996376\times 10^{-2} & 3.5788991032583934219\times 10^{-6} & 2.666889209\times 10^{-17} & 1.718004154\times 10^{-19} \\
 40 & 2.4364201607676976872\times 10^{-2} & 2.2730939533525315130\times 10^{-6} & 1.759954961\times 10^{-17} & 9.349372617\times 10^{-20} \\
 45 & 2.1720433941951952235\times 10^{-2} & 1.5205343987439192133\times 10^{-6} & 1.220841693\times 10^{-17} & 5.478475527\times 10^{-20} \\
 50 & 1.9593977138759696222\times 10^{-2} & 1.0599790877771606396\times 10^{-6} & 8.808541368\times 10^{-18} & 3.400305245\times 10^{-20} \\
 55 & 1.7846580120368287542\times 10^{-2} & 7.6418367844689967876\times 10^{-7} & 6.560715313\times 10^{-18} & 2.210127995\times 10^{-20} \\
 60 & 1.6385213628045196549\times 10^{-2} & 5.6652220737162069625\times 10^{-7} & 5.016155327\times 10^{-18} & 1.492031137\times 10^{-20} \\
 65 & 1.5144980956649394828\times 10^{-2} & 4.2998668462990765279\times 10^{-7} & 3.920290881\times 10^{-18} & 1.039710466\times 10^{-20} \\
 70 & 1.4079232946219360584\times 10^{-2} & 3.3298551248965351373\times 10^{-7} & 3.121481287\times 10^{-18} & 7.443015862\times 10^{-21} \\
 75 & 1.3153576784133680096\times 10^{-2} & 2.6238746624051624458\times 10^{-7} & 2.525601325\times 10^{-18} & 5.453231584\times 10^{-21} \\
 80 & 1.2342099554180064613\times 10^{-2} & 2.0991759354724383914\times 10^{-7} & 2.072137298\times 10^{-18} & 4.076764659\times 10^{-21} \\
 85 & 1.1624906883132949330\times 10^{-2} & 1.7019819208348517223\times 10^{-7} & 1.720986920\times 10^{-18} & 3.102161030\times 10^{-21} \\
 90 & 1.0986471753831045984\times 10^{-2} & 1.3963904283654329585\times 10^{-7} & 1.444862734\times 10^{-18} & 2.397815109\times 10^{-21} \\
 95 & 1.0414498580229646206\times 10^{-2} & 1.1578567408369181383\times 10^{-7} & 1.224760538\times 10^{-18} & 1.879447870\times 10^{-21} \\
 100 & 9.8991242481267771860\times 10^{-3} & 9.6924289089700379549\times 10^{-8} & 1.047169541\times 10^{-18} & 1.491706402\times 10^{-21} \\\end{array}$
\caption{\label{tab:redshift} Numerical results for the non-spin ($z_{(1)}$) and the linear-in-spin ($z_{(2, \chi)}$) contributions to the redshift invariant. The error on $z_{(1)}$ and $z_{(2, \chi)}$ are quantified by $ \Delta_{z}^{(1)}$ and $ \Delta_{z}^{(2, \chi)}$ respectively, which correspond to the error introduced for truncating the mode-sum at $\ell_{\mathrm{max}}$. In our calculations the truncation error was always greater than the numerical error --- we summed up to $\ell_{\mathrm{max}}=40$ and included the first seven regularisation parameters.
} 
\end{table*}

\begin{figure}[htb]
\includegraphics[width=0.95\columnwidth]{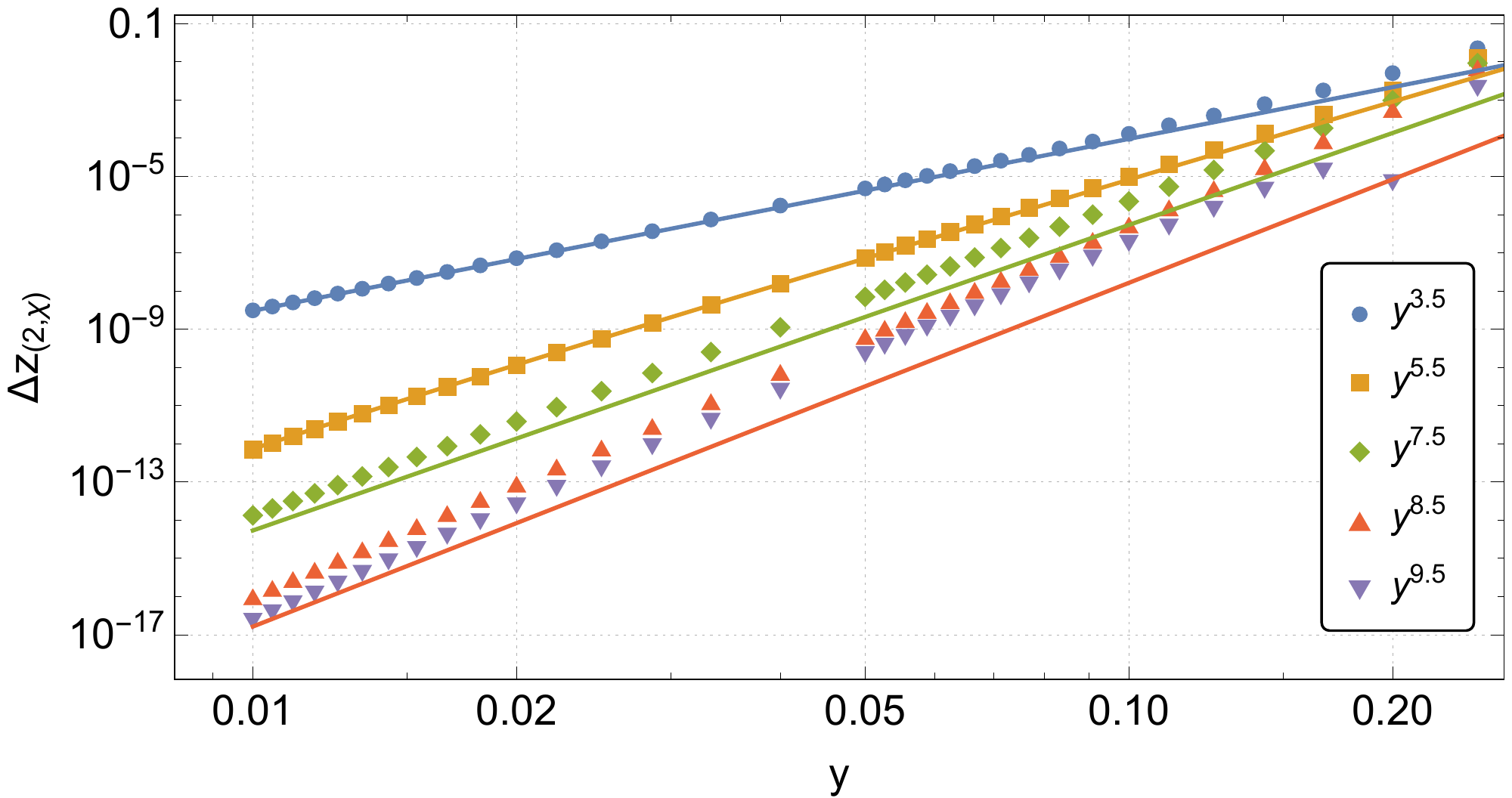}
\caption{The absolute error $\Delta z_{(2,\chi)}\equiv \left|z_{(2, \chi)}-z_{(2, \chi)}^{\mathrm{PN}} \right|$ between our numerical results and the PN series on a logarithmic scale as a function of $y$. Each set of plot markers show the comparison including increasingly high order PN terms, with the highest included order in $y$ labelled in the plot legend. The joined lines give the individual $y^{5.5}$, $y^{7.5}$, $y^{8.5}$ and $y^{9.5}$ terms in the PN series; towards the weak field these approach the leading contributions to the residuals, $\Delta z_{(2,\chi)}$, when the PN series is truncated to the preceding order.}
\label{fig:pncomparison}
\end{figure}

Strong field values of $z_{(1)}$ were computed in Refs.~\cite{Dolan:2014pja, PhysRevD.99.124046} (although both give the values for $-(u^t)^2 z_{(1)}$ and truncate the mode-sum at a larger $\ell_{\mathrm{max}}$); these agree with our non-spinning result to within the errors given. Strong field values of the linear in spin contribution to the redshift, $z_{(2,\chi)}$, have not been previously computed, although a post-Newtonian expansion was produced in Ref.~\cite{Bini:2018zde}. In Fig.~\ref{fig:pncomparison}, we compare our numerical results for $z_{(2, \chi)}$ against the equivalent post-Newtonian series from Ref.~\cite{Bini:2018zde}. As expected, the absolute error between our fully relativistic numerical results and the PN expanded redshift is higher for large values of $y$ (in the strong field) where the PN expansion breaks down. The error improves significantly further towards the weak field and improves again when adding higher order terms in the PN expansion. In the weak field, the leading contribution to the residual goes as the next (unknown) PN term beyond the order where the series was truncated.

\subsection{Waveforms}
\label{sec:waveforms}

To produce a gravitational waveform incorporating effects from a spinning secondary, we follow the two-timescale evolution and waveform-generation procedure summarised in Sec.~\ref{sec:SummaryTT}. On a grid of $\Omega$ values, we solve the RWZ equations for the master functions and calculate $C_{\ell m \omega}^{\pm}(\Omega)$ in doing so. From the $C_{\ell m \omega}^{\pm}$ values we obtain the asymptotic energy fluxes via Eq.~\eqref{eqn:Eflux} (summing to some $\ell_{\rm max}$ determined by the accuracy requirements) and in turn produce the forcing functions, $F_{\Omega}^{(0)}(\Omega)$ and $F_{\Omega}^{(1, \chi)}(\Omega)$ via Eqs.~\eqref{eq:FOmega0} and \eqref{F1spin}. We then fix the value of $\mu \chi$ and solve for the orbital phase $\phi_{p}(t)$ and frequency $\Omega(t)$.

To obtain a waveform, we note that the gravitational wave strain in the RWZ formalism is given by \cite{pound2021black}
\begin{equation}
\label{eq:waveform}
r\left(h_{+}-i h_{\times}\right)=\sum_{\ell=2}^{\infty} \sum_{m=-\ell}^{\ell} \frac{D}{2}\left(\psi_{\ell m}^{\mathrm{even}}-i \psi_{\ell m}^{\mathrm{odd}}\right)_{-2} Y_{\ell m}(\theta, \phi),
\end{equation}
where the equality holds in the limit $r \rightarrow \infty$ and the constant $D\equiv\sqrt{(\ell-1) \ell(\ell+1)(\ell+2)}$. The function $_{-2} Y_{\ell m}(\theta, \phi)$ is the spin-weight $s=-2$ spherical harmonic. Conveniently, $\psi_{\ell m}^{\mathrm{even/odd}} \rightarrow C_{\ell m \omega}^{+ \mathrm{even/odd}}e^{-i m \phi_{p}(t)}$ as $r \rightarrow \infty$, so defining $h \equiv r\left(h_{+}-i h_{\times}\right)$ and explicitly taking the limit in Eq.~\eqref{eq:waveform} we can write
\begin{equation}
\label{eq:waveformamps}
h=\sum_{\ell=2}^{\infty} \sum_{m=-\ell}^{\ell} h_{\ell m}(t) \,_{-2} Y_{\ell m}(\theta, \phi),
\end{equation}
where the spherical harmonic modes of the waveform are given by
\begin{align}
h_{\ell m} (t) &= 
\label{eq:hlm}
\begin{dcases}
\frac{D}{2} C_{\ell m \omega}^{+}\left(\Omega(t)\right) e^{-i m \phi_{p}(t)} & \ell+m \quad \text{even},\\
-\frac{ i D}{2} C_{\ell m \omega}^{+}\left(\Omega(t)\right) e^{-i m \phi_{p}(t)} & \ell+m \quad \text{odd}.
 \end{dcases} 
\end{align}

\begin{figure}[htb]
\includegraphics[width=0.95\columnwidth]{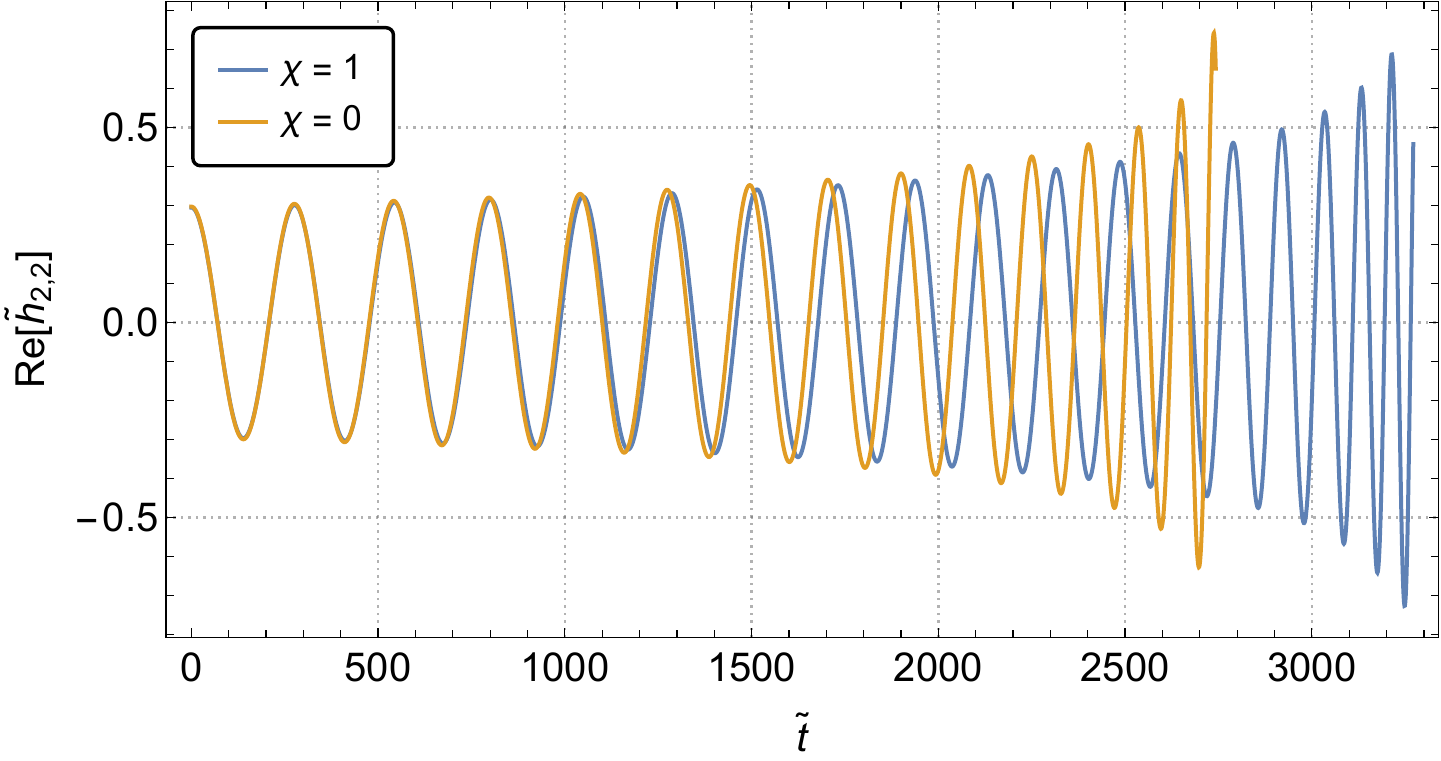}
\caption{Comparison of the $(\ell,m)=(2,2)$ mode of the waveform with $\chi=1$ (blue) and $\chi=0$ (orange) for an equal mass binary. Both inspirals begin at the same orbital frequency ($r_\Omega=20M$) and phase before evolving.}
\label{fig:waveforma}
\end{figure}

\begin{figure}[htb]
\includegraphics[width=\columnwidth]{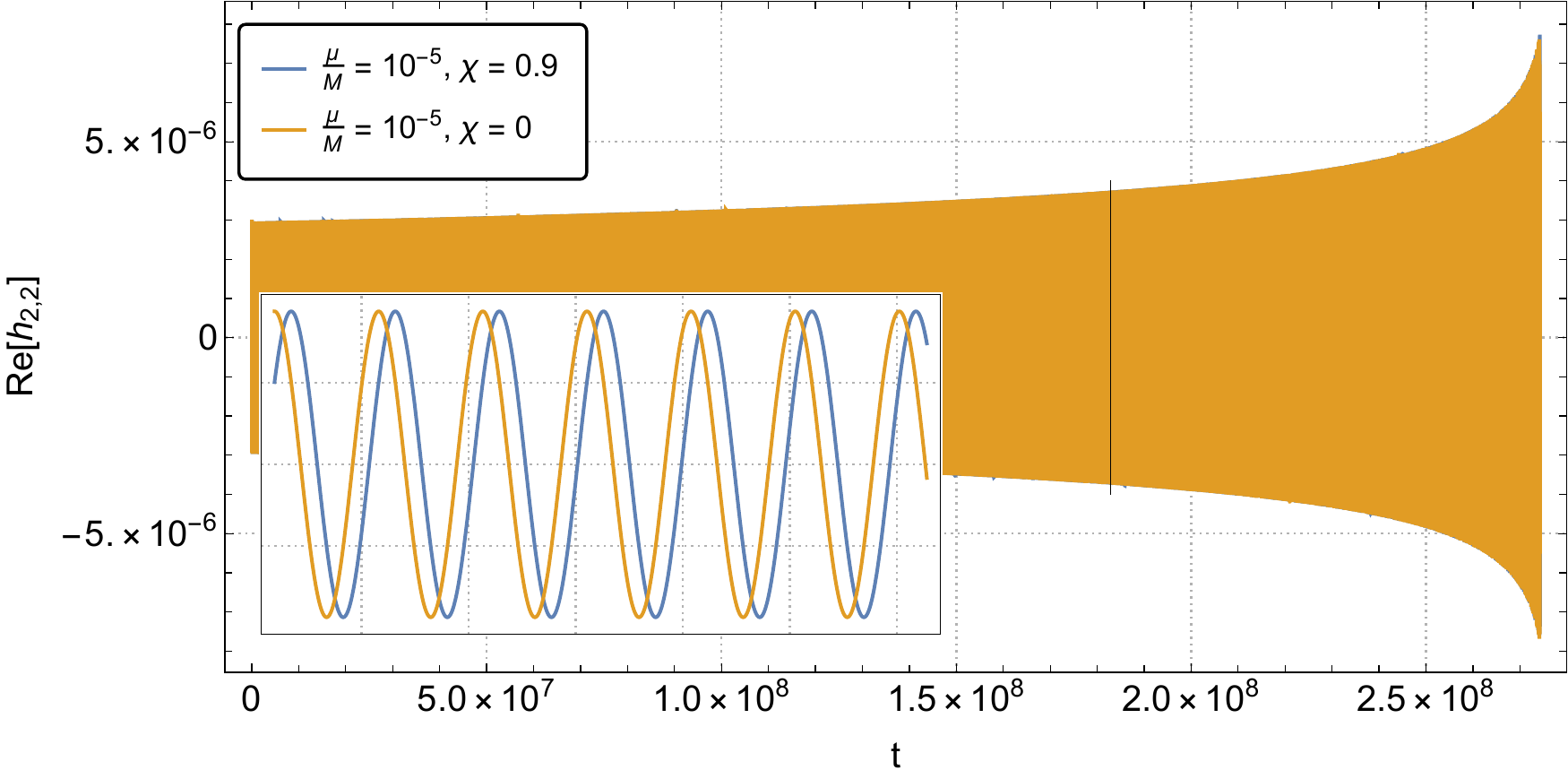}
\caption{Comparison as in Fig.~\ref{fig:waveforma} but for a binary with a more realistic EMRI mass ratio of $1:10^{5}$. }
\label{fig:waveformb}
\end{figure}

\begin{figure}[htb]
\includegraphics[width=0.95\columnwidth]{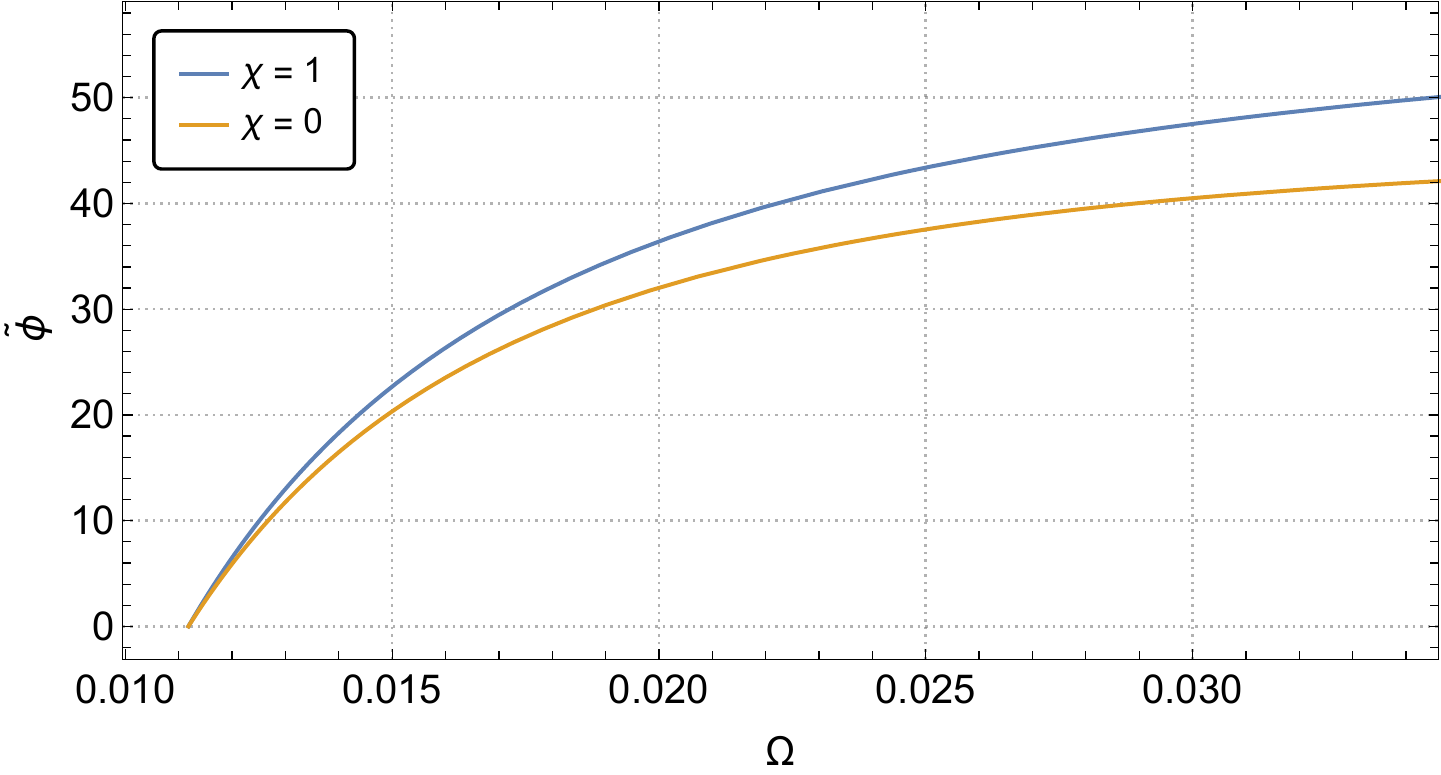}
\caption{Comparison of the waveform phase, $\tilde{\phi}$, as a function of orbital frequency with $\chi=1$ (blue) and $\chi=0$ (orange) for an equal mass binary.}
\label{fig:phase}
\end{figure}

\begin{figure}[htb]
\includegraphics[width=0.95\columnwidth]{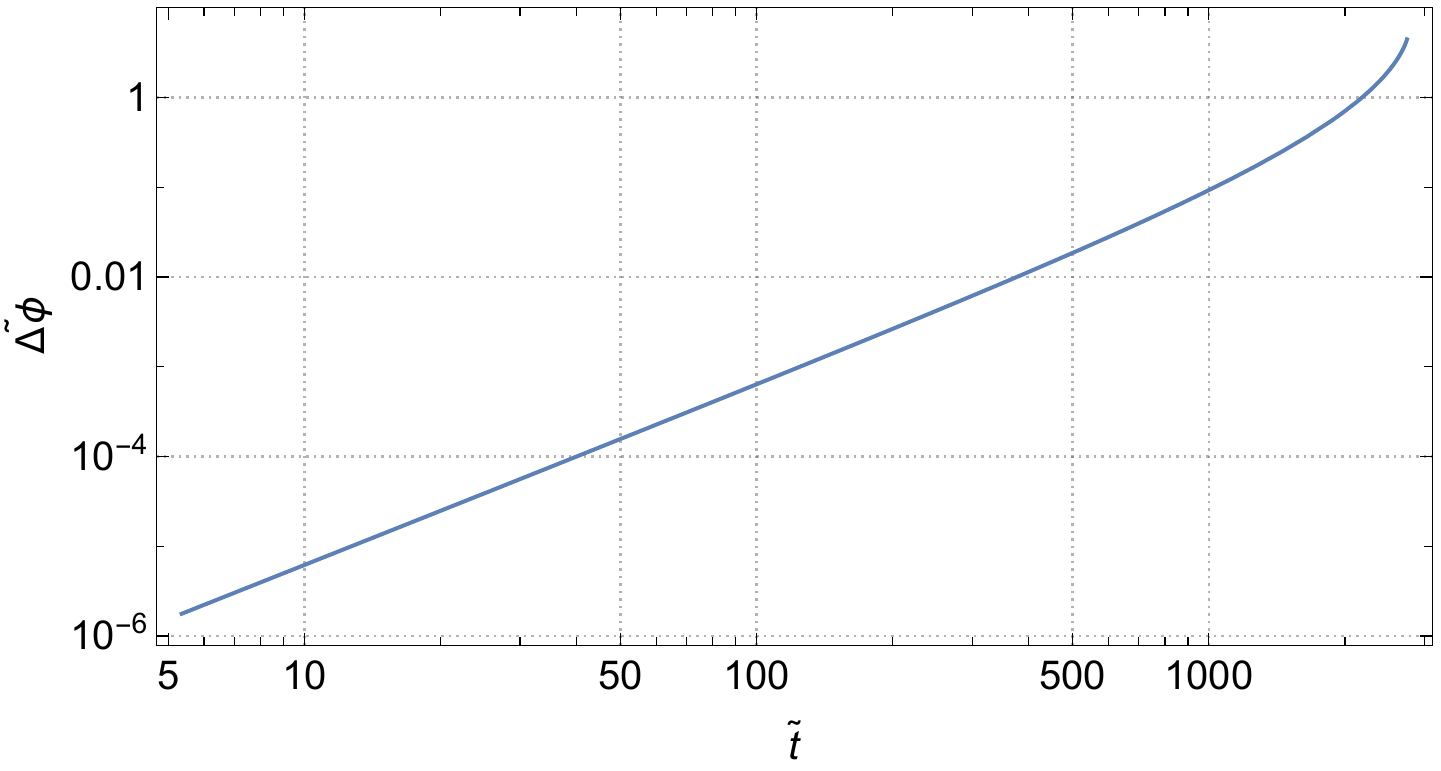}
\caption{Phase difference (in radians) between the waveform with a spinning seocondary and with a non-spinning secondary, $\Delta \tilde{\phi} = \tilde{\phi}_0 -  \tilde{\phi}$.}
\label{fig:dephasing}
\end{figure}

In Figures \ref{fig:waveforma} and \ref{fig:waveformb} we plot the $(\ell,m)=(2,2)$ mode of the waveform for mass ratio $1:1$ and $1:10^5$ binaries, respectively. Figures \ref{fig:phase} and \ref{fig:dephasing} demonstrate the impact of the spin on the waveform's phase. The leading phase difference between a waveform including the secondary's spin and a non-spinning waveform comes from the linear in spin contribution to $\tilde{\Delta \phi}=\tilde{\phi}_0 -  \tilde{\phi} \sim - \e \chi \tilde{\phi}_\chi = - \chi \phi_\chi.$ Thus the leading phase difference is independent of the mass ratio and Fig.~\ref{fig:dephasing} can be rescaled for any mass ratio and $\chi$ by simply multiplying $\tilde{\phi}$ by $\chi$ and $\tilde{t}$ by $\e^{-1}$, giving the error to expect on the waveform's phase accumulated over time if neglecting the spin of the secondary. This highlights the importance of including the secondary's spin in waveform models, as a key requirement to test fundamental physics with EMRI waveforms is to accurately track the phase over the full inspiral.

\section{Conclusions}
\label{sec:conclusions}

We have performed calculations of the leading order self-force including the sub-leading linear-in-spin effects in the two-timescale expansion and demonstrated how these spin effects can be included in 1PA waveforms. In doing so we have produced waveforms complete at adiabatic order and including the spin's 1PA contribution. We have developed a fixed frequency approach for solving the perturbation equations which significantly improves the computational efficiency of linear-in-spin self-force quantities, including the asymptotic gravitational wave fluxes and the metric perturbation. We provided the first fully relativistic regularisation scheme to treat the singular field associated with a spinning secondary body, using a first principles approach --- having derived the spin-dipole contributions to the Detweiler-Whiting singular field and produced a covariant expansion of the approximated field. We have performed the first fully relativistic calculations of a conservative self-force quantity with a spinning secondary --- having computed the redshift invariant and checked agreement with the equivalent PN expansion.

There are several obvious future extensions to this work. Firstly, we could repeat our calculations for more general EMRI configurations, allowing for eccentric equatorial or generic motion of the secondary body and performing the computations in a Kerr background spacetime to include the spin of the \textit{primary}.  Further generalization beyond equatorial motion requires treatment of the secondary body's precessing spin (instead of the spin aligned/anti-aligned case). Now that the regularisation scheme has been developed for a spinning body, we could extend current calculations of other gauge invariants such as the spin precession invariant to include the spin of the secondary. We could also study more extended body effects, beginning with including the secondary's quadrupole moment in our computations. Finally, as progress on 2SF calculations continue, it would be natural to compute complete 1PA waveforms, including all the conservative 1SF and dissipative 2SF effects as well as the spin effects highlighted in this work. Aside from their immediate application for LISA, complete 1PA waveforms could be compared with numerical relativity counterparts and used to establish the domain of validity of perturbation theory in less extreme mass ratios.

\begin{acknowledgments}

The authors thank Gabriel Piovano, Vojtech Witzany and Chris Kavanagh for helpful discussions.
JM acknowledges support from the Irish Research Council under grant GOIPG/2018/448. AP acknowledges the support of a Royal Society University Research Fellowship. This work makes use of the Black Hole Perturbation Toolkit \cite{BHPToolkit} and xAct \cite{xact}.

\end{acknowledgments}

\appendix

\begin{widetext}
\section{Stress-energy tensor for a point mass with aligned spin on a circular orbit}
\label{sec:K}

The coefficients appearing in Eq.~\eqref{eq:Tmunu} are given explicitly by
\begin{align}
&K_{0}^{t t}=u^{t}, \quad K_{0}^{t \phi}=u^{\phi}, \quad K_{0}^{\phi \phi}=\left(u^{\phi}\right)^{2} / u^{t}, \nonumber \\
&K_{1}^{t t}=\frac{- M^{3 / 2}}{r_{\Omega}\left(r_{\Omega}-2 M\right) \sqrt{r_{\Omega}-3 M}}, \quad K_{1}^{t \phi}=K_{1}^{\phi t}=-\frac{M}{r_{\Omega}^{5 / 2} \sqrt{r_{\Omega}-3 M}}, \nonumber \\
&K_{1}^{r r}=-\frac{M^{1 / 2}\left(r_{\Omega}-2 M\right) \sqrt{r_{\Omega}-3 M}}{r_{\Omega}^{3}}, \quad K_{1}^{\phi \phi}=-\frac{M^{1 / 2}\left(r_{\Omega}-2 M\right)}{r_{\Omega}^{4} \sqrt{r_{0}-3 M}}, \nonumber \\
&K_{2}^{t r}=K_{2}^{r t}=\frac{\sqrt{r_{\Omega}-3 M}}{2 r_{\Omega}^{3 / 2}}, \quad K_{2}^{r \phi}=K_{2}^{\phi r}=\frac{M^{1/2}\sqrt{r_{\Omega}-3 M}}{2 r_{\Omega}^{3}}, \nonumber \\
&K_{3}^{t t}=-\frac{M^{1/2}}{\sqrt{r_{\Omega}-3 M}}, \quad K_{3}^{t \phi}=K_{3}^{\phi t}=-\frac{(r_{\Omega}-M)}{2 r_{\Omega}^{3 / 2} \sqrt{r_{\Omega}-3 M}}, \quad K_{3}^{\phi \phi}=-\frac{M^{1/2}\left(r_{\Omega}-2 M\right)}{r_{\Omega}^{3} \sqrt{r_{\Omega}-3 M}}.
\end{align}

\section{Sources for the RWZ equations at fixed frequency}
\label{sec:sources}

\subsection{Odd parity}
\label{sec:oddapp}

The Cunningham-Price-Moncrief master function is defined in \cite{PhysRevD.71.104003} and satisfies the RWZ equation with the Regge-Wheeler potential;
\begin{equation}
V_{\text{odd}}\equiv \frac{f}{r^2}\left[\ell(\ell+1) - \frac{6M}{r} \right].
\end{equation}

For a spin-aligned secondary in a circular equatorial orbit parameterised at fixed frequency, the frequency domain source \eqref{eqn:explicitsource} in the odd parity sector has the following coefficients of the distributional functions:
\begin{align*}
H^{(2, \chi)}_{\ell m \omega}&=  -\frac{4 \pi \mu^2 \chi f_\Omega^2}{\lambda (\lambda+1) } \sqrt{1 - \frac{3M}{r_\Omega}}\sqrt{(\ell-m)(1+\ell+m)} Y_{\ell (m+1)}\left(\frac{\pi}{2},0\right), \\
F^{(1)}_{\ell m \omega}&= \frac{8 \pi \mu f_\Omega^2}{\lambda (\lambda+1) }\sqrt{\frac{M}{r_\Omega - 3M}}\sqrt{(\ell-m)(1+\ell+m)} Y_{\ell (m+1)}\left(\frac{\pi}{2},0\right), \\
F^{(2, \chi)}_{\ell m \omega}&= -\frac{8 \pi \mu^2 \chi f_\Omega}{\lambda (\lambda+1)}\frac{\left( M^2 +3M r_\Omega -r_\Omega^2\right)}{r_\Omega^3}\sqrt{\frac{r_\Omega}{r_\Omega - 3M}}\sqrt{(\ell-m)(1+\ell+m)} Y_{\ell (m+1)}\left(\frac{\pi}{2},0\right), \\
G^{(1)}_{\ell m \omega}&=- \frac{8 \pi \mu f_\Omega}{\lambda (\lambda+1) r_\Omega}\sqrt{\frac{M}{r_\Omega - 3M}}\sqrt{(\ell-m)(1+\ell+m)} Y_{\ell (m+1)}\left(\frac{\pi}{2},0\right), \\
G^{(2, \chi)}_{\ell m \omega}&= \frac{4 \pi \mu^2 \chi M \left(m^2 r_\Omega +(2-3m^2)M\right)}{\lambda (\lambda+1) r_\Omega^4}\sqrt{\frac{r_\Omega}{r_\Omega - 3M}}\sqrt{(\ell-m)(1+\ell+m)} Y_{\ell (m+1)}\left(\frac{\pi}{2},0\right),
\end{align*}
where $\lambda \equiv (\ell-1)(\ell+2)/2$ and we have separated the spin independent and linear-in-spin contributions as $\bar{F}_{\ell m \omega}\equiv F^{(1)}_{\ell m \omega}+ F^{(2, \chi)}_{\ell m \omega}$, $\bar{G}_{\ell m \omega}\equiv G^{(1)}_{\ell m \omega}+  G^{(2, \chi)}_{\ell m \omega}$ and $\bar{H}_{\ell m \omega}\equiv H^{(2, \chi)}_{\ell m \omega}$.
\subsection{Even parity}
\label{sec:evenapp}

The Zerilli-Moncrief master function is defined in \cite{PhysRevD.71.104003} and satisfies the RWZ equation with the Zerilli potential;

\begin{equation}
V_{\text{even}}\equiv \frac{f}{r^2 \Lambda^{2}}\left[2\lambda^{2} \left(\lambda+1+\frac{3M}{r} \right) + \frac{18M^{2}}{r^{2}} \left(\lambda+\frac{M}{r} \right) \right].
\end{equation}

For a spin-aligned secondary in a circular equatorial orbit parameterised at fixed frequency, the frequency domain source \eqref{eqn:explicitsource} in the even parity sector has the following coefficients of the distributional functions:
\begin{align*}
H^{(2, \chi)}_{\ell m \omega} = & 0, \\
F^{(1)}_{\ell m \omega}= & \frac{8 \pi \mu r_\Omega f_\Omega^3}{(\lambda r_\Omega +3M) (\lambda+1) }\sqrt{\frac{r_\Omega}{r_\Omega - 3M}} Y_{\ell m}\left(\frac{\pi}{2},0\right), \\
F^{(2, \chi)}_{\ell m \omega}= &-\frac{8 \pi \mu^2 \chi f_\Omega (r_\Omega-3 M) }{(\lambda r_\Omega +3M) (\lambda+1) r_\Omega^2} \sqrt{\frac{M}{r_\Omega - 3M}} Y_{\ell m}\left(\frac{\pi}{2},0\right)\left( 3 \left(1-m^2\right) \frac{M}{\lambda}+ r_\Omega \left(\lambda -m^2\right)+\frac{\left(-7M^2-M r_\Omega+r_\Omega^2\right)}{(r_\Omega-3 M) } \right), \\
G^{(1)}_{\ell m \omega}= &\frac{8 \pi  \mu  f_\Omega}{ (\lambda +1) r_\Omega^2
   (3 M+\lambda  r_\Omega)^2} \sqrt{\frac{r_\Omega}{r_\Omega-3M}}Y_{\ell m}\left(\frac{\pi}{2},0\right) \left(3 M^3 \left(5  +(3/\lambda)(m^2-1)\right)+2  M^2 r_\Omega \left(2
   \lambda +3 m^2-9\right) \right. \\
   &\qquad \qquad \qquad \qquad  \qquad \qquad \qquad  \qquad \qquad \qquad \qquad \qquad \qquad \qquad \qquad \left. +\lambda M r_\Omega^2 \left(\lambda +m^2-4\right)-\lambda (\lambda +1) r_\Omega^3\right), \\
G^{(2, \chi)}_{\ell m \omega}= & -\frac{8 \pi  \mu ^2 \chi }{\lambda  (\lambda +1) r_\Omega^4 (3 M+\lambda  r_\Omega)^2} \sqrt{\frac{M}{r_\Omega-3 M}} Y_{\ell m}\left(\frac{\pi}{2},0\right)\left(3 \lambda  m^2 M r_\Omega \left(2 M^2+3 M r_\Omega-r_\Omega^2\right) \right. \\
&\qquad\qquad \qquad\qquad  \qquad\qquad \qquad  \left.+(4 M-r_\Omega) \left(9 \left(m^2-1\right) M^3-3 \lambda  M \left(M^2+M r_\Omega+r_\Omega^2\right) +\lambda^3 r_\Omega^2 (M-r_\Omega)\right) \right. \\
& \qquad\qquad \qquad \qquad\qquad \qquad\qquad \quad \left. +\lambda ^2 r_\Omega \left(-m^2 r_\Omega \left(2 M^2-4 M r_\Omega+r_\Omega^2\right)+4 M^3-10 M^2 r_\Omega-2 M r_\Omega^2+r_\Omega^3\right)\right).
\end{align*}

\section{Fixed-radius approaches to numerical RWZ solutions}
\label{sec:linearisenumerics}

Earlier works that parameterise the secondary body at fixed radius extract out the linear in spin terms from numerical results by repeating the calculations for multiple values of $\chi$ and fitting a polynomial to the results \cite{Akcay:2019bvk, Piovano:2020zin}. This method is undesirable as it is very computationally expensive. It is also possible to linearize the perturbation equations themselves and solve a coupled system with an extended source (as in Ref.~\cite{Piovano:2021iwv}). That way, numerical results relying on the integration of the perturbation equations may be obtained directly to linear order in spin. In the RWZ formalism, the linearized perturbation equations in the frequency domain are
\begin{align}
&\mathcal{L}_0 \psi^{0}_{\ell m \omega}(r)=Z^{0}_{\ell m \omega}(r), \label{eqn:RWZnospin} \\
\label{eqn:RWZlinearspin}
&\mathcal{L}_0 \psi^{\chi}_{\ell m \omega}(r)=Z^{\chi}_{\ell m \omega}(r) -2 \omega_{0} \omega_{\chi} \psi^{0}_{\ell m \omega}(r),
\end{align}
where we have defined the RWZ `non-spin' operator as:
\begin{equation*}
\mathcal{L}_0\equiv \left[\dfrac{\partial^{2}}{\partial r_{*}^{2}} -V_{\ell}(r) + \omega_0^{2} \right],
\end{equation*}
and separated the spin independent and linear-in-spin terms as $\psi_{\ell m \omega}(r)=\psi^{0}_{\ell m \omega}(r)+\psi^{\chi}_{\ell m \omega}(r)$, $Z_{\ell m \omega}(r)=Z^{0}_{\ell m \omega}(r)+Z^{\chi}_{\ell m \omega}(r)$ and $\omega=\omega_0+\omega_\chi$. Eq.~\eqref{eqn:RWZnospin} is the usual RWZ equation for a non-spinning secondary body. Eq.~\eqref{eqn:RWZlinearspin} for the linear in spin master function has the same differential operator as Eq.~\eqref{eqn:RWZnospin} but sourced by both the compact spin source $Z^{\chi}_{\ell m \omega}(r)$ and an extended source term from the coupling to $\psi^{0}_{\ell m \omega}(r)$.

In general, we wish to avoid modelling with an extended source. This becomes especially important when generalising to eccentric (or more generic) motion of the secondary and using the method of extended homogeneous solutions (EHS) to solve the perturbation equations \cite{Barack:2008ms, PhysRevD.82.084010} instead of regular variation of parameters. Motivated by this, we split the extended and compact parts of Eq.~\eqref{eqn:RWZnospin}  into two separate equations --- by linearity we can write $\psi^{\chi}_{\ell m \omega}(r)=\psi^{\chi, \rm ext}_{\ell m \omega}(r)+\psi^{\chi, \rm c}_{\ell m \omega}(r)$.  Acting on the extended equation again, having identified the RWZ operator as a `partial annihilator' \cite{Hopper:2012ty}, we obtain
\begin{align}
&\mathcal{L}_0 \psi^{\chi, \rm c}_{\ell m \omega}(r)=Z^{\chi}_{\ell m \omega}(r), \label{eqn:RWZspincompact}\\
&\mathcal{L}_0^2 \psi^{\chi, \rm ext}_{\ell m \omega}(r)=-2 \omega_{0} \omega_{\chi} Z^{0}_{\ell m \omega}(r), \label{eqn:RWZspinext}
\end{align}
where now the entire system involves only compact sources and we have decoupled the equations. The trade off is having to solve the fourth order Eq.~\eqref{eqn:RWZspinext} which has four homogeneous solutions to find numerically. Conveniently, only two new homogeneous solutions are required in practice as the other two are shared with Eq.~\eqref{eqn:RWZnospin} --- the `in' and `up' solutions. The two new homogeneous solutions to Eq.~\eqref{eqn:RWZspinext} can be considered the extended `in' and `up' solutions and the boundary conditions are similarly posed by considering their asymptotic behaviour
\begin{align}
\label{bcs2}
&\hat{R}_{\ell m\omega}^{\chi, \rm ext -}(r_* \rightarrow -\infty) \sim r_* e^{-i \omega_0 r_*},\\
&\hat{R}_{\ell m\omega}^{\chi, \rm ext +}(r_* \rightarrow \infty) \sim r_* e^{i \omega_0 r_*}.
\end{align}
Once all the homogeneous solutions are obtained numerically, the general solutions for the spin-linearized master functions are easily obtained via standard variation of parameters. The method can be extended to eccentric orbits using EHS.

\textit{However, the issue of linearizing numerical quantities such as the master functions is entirely avoided with the fixed frequency parametrisation} where $\omega_\chi=0$. Then, the homogeneous RWZ equation is independent of spin and may be solved in the same way as for a non-spinning secondary. The linearized inhomogeneous solution for the retarded master function is obtained analytically from the homogeneous solutions as described in Section \ref{sec:retsoln} in Eqs.~(\ref{eqn:formR}, \ref{MatchingCalc}) --- in practice this is equivalent to solving Eqs.~\eqref{eqn:RWZnospin} and \eqref{eqn:RWZspincompact} with $\psi^{\chi, \rm ext}_{\ell m \omega}(r)=0$. 

\section{Derivation of the spin-dipole singular field}
\label{sec:singular-dipole}

Substituting the stress-energy in Eq.\eqref{stressenergyspin} and the singular Green function in Eq.~\eqref{hadamard} into Eq.~\eqref{lorenzpert},
\begin{align}
\bar{h}_{\alpha \beta}^{\rm S (\chi)}(x)& = 4 \int G^{\rm S}_{\alpha \beta \alpha'\beta'}(x,x')\nabla_{\rho'} \left( \int d \tau'  \, \frac{\delta^{4}\left[x'-z(\tau' )\right]}{\sqrt{-g'}} u^{(\alpha'}\tilde{S}^{\beta') \rho'}\right)\sqrt{-g'}d^4x' \nonumber \\
& = -2 \int \delta[\sigma] \left[ \nabla_{\rho'}  U_{\alpha \beta \alpha' \beta'} + V_{\alpha \beta \alpha' \beta'} \sigma_{\rho'}\right] u^{(\alpha'}\tilde{S}^{\beta') \rho'} d \tau' -2 \int   \delta'\left[\sigma \right] U_{\alpha \beta \alpha' \beta'}  \sigma_{\rho'} u^{(\alpha'}\tilde{S}^{\beta') \rho'} d \tau' \nonumber \\
&  \quad -2 \int   \theta\left[\sigma \right] \nabla_{\rho'}  V_{\alpha \beta \alpha' \beta'} u^{(\alpha'}\tilde{S}^{\beta') \rho'}  d \tau',
\end{align} 
where in this context $\delta[\sigma]= \delta[\sigma(x, z(\tau'))]$.
Considering the three different distributional integrals separately, the first integral is
\begin{gather}
 -2  \int  \left( \frac{\delta\left[\tau'-\tau_{A}\right]}{\left|\sigma_{c'}u^{c'}\right|} + \frac{\delta\left[\tau'-\tau_{R}\right]}{\left|\sigma_{c'}u^{c'}\right|}   \right) \left( \nabla_{\rho'}  U_{\alpha \beta a^{\prime} b^{\prime}} + V_{\alpha \beta a^{\prime} b^{\prime}} \sigma_{\rho'}\right)u^{(\alpha'}\tilde{S}^{b') \rho'}  d \tau' \nonumber \\
  = -2   \left. \left[ \frac{\left( \nabla_{\rho'}  U_{\alpha \beta a^{\prime} b^{\prime}} + V_{\alpha \beta a^{\prime} b^{\prime}} \sigma_{\rho'}\right)u^{(\alpha'}\tilde{S}^{b') \rho'}}{\left|\sigma_{c'}u^{c'}\right|} \right] \right|_{x'=x_{A/R}},
 \end{gather}
where $\delta\left[\sigma\right]$ was rewritten using equation~\eqref{deltacomposition}, the fact that the two simple roots of $\sigma\left(x, z(\tau')\right)$ along the worldline are $x_{A}/x_{R}$, and that $\frac{d}{d\tau'}\sigma(x,z(\tau'))=\sigma_{c'}u^{c'}$. The second integral is
\begin{gather}
   -2  \int \left(  \frac{\delta'\left[\tau'-\tau_{A}\right]+\delta'\left[\tau'-\tau_{R}\right]}{\left|\sigma_{c'}u^{c'}\right|\sigma_{d'}u^{d'}} + \frac{\left( \delta\left[\tau'-\tau_{A}\right] +\delta\left[\tau'-\tau_{A}\right]\right)\sigma_{p' q'}u^{p'}u^{q'}}{\left|\sigma_{c'}u^{c'}\right|(\sigma_{d'}u^{d'})^2} \right) U_{\alpha \beta a^{\prime} b^{\prime}}  \sigma_{\rho'}u^{(\alpha'}\tilde{S}^{b') \rho'}  d \tau'\nonumber \\ 
  = 2 \left. \left[ \left( \frac{u^{p'}\nabla_{p'}U_{\alpha \beta a^{\prime} b^{\prime}}  \sigma_{\rho'} +U_{\alpha \beta a^{\prime} b^{\prime}}  u^{q'}\sigma_{\rho' q'}}{\left|\sigma_{c'}u^{c'}\right|\sigma_{d'}u^{d'}} -  \frac{U_{\alpha \beta a^{\prime} b^{\prime}} \sigma_{\rho'}\sigma_{p' q'}u^{p'}u^{q'}}{\left|\sigma_{c'}u^{c'}\right|(\sigma_{d'}u^{d'})^2} \right)u^{(\alpha'}\tilde{S}^{b') \rho'} \right] \right|_{x'=x_{A/R}},
\end{gather}
where the $\delta'\left[\sigma \right]$ has been rewritten using equation~\eqref{deltaprimecomp}. Also, although strictly $\frac{d^2}{d\tau'^2}\sigma(x,z(\tau'))=\sigma_{c' d'}u^{c'}u^{d'}+\frac{du^{c'}}{d\tau'}\sigma_{c'}$, Eq.~\eqref{eqn:MPD} gives that $\frac{du^{c'}}{d\tau'}=O(\e)$ and thus four-acceleration can be neglected in terms that are already linear order in spin. Similarly, Eq.~\eqref{eqn:MPD2} implies that $\tilde{S}^{\alpha \beta}$ may be treated as a constant to first order in spin.
Combining everything, the spin-dipole contribution to the Detweiler-Whiting singular field is
\begin{align}
\bar{h}_{\alpha\beta}^{\rm S (\chi)}(x) &= 2 \Bigg[ \bigg( \frac{u^{\rho'}\nabla_{\rho'}U_{\alpha\beta\alpha'\beta'}  \sigma_{\rho'}+U_{\alpha\beta\alpha'\beta'}  u^{\kappa'}\sigma_{\rho' \kappa'}}{\sigma_{\delta'}u^{\delta'}}
 - ( \nabla_{\rho'}  U_{\alpha\beta\alpha'\beta'} + V_{\alpha\beta\alpha'\beta'} \sigma_{\rho'})\nonumber \\
& \qquad
 -  \frac{U_{\alpha\beta\alpha'\beta'} \sigma_{\rho'}\sigma_{\rho' \kappa'}u^{\rho'}u^{\kappa'}}{(\sigma_{\delta'}u^{\delta'})^2}\bigg)  \frac{u^{(\alpha'}\tilde{S}^{\beta') \rho'}}{|\sigma_{\gamma'}u^{\gamma'}|} \Bigg] \Bigg|_{x'=x_{A/R}}  -  2 \int_{\tau_{R}}^{\tau_{A}} \nabla_{\rho'}  V\left(x, z(\tau')\right)_{\alpha\beta\alpha'\beta'} u^{(\alpha'}\tilde{S}^{\beta') \rho'} d \tau',
\end{align}

\section{Covariant expansion of the Detweiler-Whiting singular field}
\label{sec:SingulFieldExpansion}

We have already given the expression for the covariant expansion of the Detweiler-Whiting singular field to linear order in spin, for a particle under the pole-dipole approximation;
\begin{equation*}
\bar{h}_{\mu \nu}^{\rm S}=\mu^2 \frac{4 g_{\mu}^{\bar{\alpha}} g_{\nu}^{\bar{\beta}} u_{(\bar{\alpha}} \tilde{S}_{\bar{\beta}) \bar{\gamma}} \sigma^{\bar{\gamma}}}{\bar{\lambda}^{2} \bar{s}^{3}}+\mu \frac{4 g_{\mu}^{\bar{\alpha}} g_{\nu}^{\bar{\beta}} u_{\bar{\alpha}} u_{\bar{\beta}}}{\bar{\lambda} \bar{s}}+\bar{h}_{\mu \nu}^{\rm S(0)}+\bar{\lambda} \bar{h}_{\mu \nu}^{\rm S(1)}+\mathcal{O}\left(\bar{\lambda}^2\right).
\end{equation*}\\

The second two terms are given explicitly by
\begin{align}
&\bar{h}_{\mu \nu}^{\rm S(0)}=-\mu^2\frac{g_{\mu}^{\bar{\alpha}} g_{\nu}^{\bar{\beta}}\tilde{S}^{\bar{\gamma} \bar{\delta}}}{3 \bar{s}^5}\left[ 3 \bar{r} \bar{s}^4 \left( 2 g_{\bar{\gamma} (\bar{\alpha}} R_{\bar{\beta})u \bar{\delta}u} + 2 u_{(\bar{\alpha}}R_{\bar{\beta}) \bar{\gamma} \bar{\delta} u} +4 u_{(\bar{\alpha}}R_{\bar{\beta}) u\bar{\gamma} \bar{\delta}} \right) \nonumber \right. \\
& \left. -(3\bar{r}^2 - \bar{s}^2)u_{(\bar{\alpha}}g_{\bar{\beta})\bar{\gamma}}\sigma_{\bar{\delta}}R_{u \sigma u \sigma} + \bar{r}\bar{s}^2 \left(2g_{\bar{\gamma} ( \bar{\alpha}}u_{\bar{\beta})}R_{\bar{\delta} \sigma u \sigma} -6g_{\bar{\gamma} ( \bar{\alpha}}R_{\bar{\beta})uu\sigma}\sigma_{\bar{\delta}}  -6u_{(\bar{\alpha}}R_{\bar{\beta}) \bar{\gamma} u \sigma}\sigma_{\bar{\delta}} \right) \nonumber \right. \\
& \left. + \bar{s}^4 \left( 6 g_{\bar{\gamma} ( \bar{\alpha}} R_{\bar{\beta}) u \bar{\delta} \sigma} -2g_{\bar{\gamma} ( \bar{\alpha}}u_{\bar{\beta})}R_{\bar{\delta} u u \sigma} + 6 u_{(\bar{\alpha}}R_{\bar{\beta} \bar{\gamma} \bar{\delta} \sigma} -6 R_{\bar{\alpha} \bar{\gamma} \bar{\beta} u}\sigma_{\bar{\delta}} -6 R_{\bar{\alpha} u \bar{\beta} \bar{\gamma}}\sigma_{\bar{\delta}} -3 u_{\bar{\alpha}}u_{\bar{\beta}}R_{\bar{\gamma} \bar{\delta} u \sigma}\right) \nonumber \right. \\
& \left. +\bar{r}^2\bar{s}^2 \left(2 u_{(\bar{\alpha}}g_{\bar{\beta}) \bar{\gamma}} R_{\bar{\delta}uu\sigma} +3u_{\bar{\alpha}}u_{\bar{\beta}}R_{\bar{\gamma} \bar{\delta} u \sigma} \right) \right],
\end{align}
and
\begin{align}
&\bar{h}_{\mu \nu}^{\rm S(1)}= \mu \frac{2 g_{\mu}^{\bar{\alpha}} g_{\nu}^{\bar{\beta}}}{3 \bar{s}^{3}}\left[\left(\bar{r}^{2}-\bar{s}^{2}\right) R_{u \sigma u \sigma} u_{\bar{\alpha}} u_{\bar{\beta}}-6 \bar{r} \bar{s}^{2} R_{u \sigma u(\bar{\alpha}} u_{\bar{\beta})}-6 \bar{s}^{4} R_{\bar{\alpha} u \bar{\beta} u}\right]\\
&+ \mu^2 \frac{g_{\mu}^{\bar{\alpha}} g_{\nu}^{\bar{\beta}}\tilde{S}^{\bar{\gamma} \bar{\delta}}}{3 \bar{s}^3}\left[\bar{s}^2(\bar{r}^2+\bar{s}^2)\left( -4g_{\bar{\gamma} ( \bar{\alpha}} R_{\bar{\beta}) u \bar{\delta} u;u} -4u_{(\bar{\alpha}}R_{\bar{\beta}) \bar{\gamma} \bar{\delta} u;u} - 6u_{(\bar{\alpha}}R_{\bar{\beta}) u\bar{\gamma} \bar{\delta} ;u} \right)  -\bar{r}^3 R_{\bar{\gamma}\bar{\delta} u \sigma ;u } u_{\bar{\alpha}}u_{\bar{\beta}} + 3\bar{s}^4(R_{\bar{\alpha} \bar{\gamma} \bar{\beta} u ; \bar{\delta}} +R_{\bar{\alpha} u \bar{\beta} \bar{\gamma} ; \bar{\delta}}) \nonumber \right. \\
&\left. +\bar{s}^2 \left(2 g_{\bar{\gamma} ( \bar{\alpha}}R_{\bar{\beta})u \bar{\delta} \sigma; \sigma} +2 u_{(\bar{\alpha}}R_{\bar{\beta}) \bar{\gamma} \bar{\delta} \sigma ; \sigma} -3R_{\bar{\alpha}\bar{\gamma} \bar{\beta} u ; \sigma} \sigma_{\bar{\delta}} -3R_{\bar{\alpha}u \bar{\beta} \bar{\gamma} ; \sigma} \sigma_{\bar{\delta}} -4 g_{\bar{\gamma} ( \bar{\alpha}}R_{\bar{\beta})u u \sigma ;u}\sigma_{\bar{\delta}} -4 u_{(\bar{\alpha}}R_{\bar{\beta}) \bar{\gamma} u \sigma ; u} \sigma_{\bar{\delta}}\right) \nonumber \right. \\
&\left. +\bar{r}\bar{s}^2 \left(2g_{\bar{\gamma}(\bar{\alpha}}(R_{\bar{\beta}) u \bar{\delta} u; \sigma}-2(R_{\bar{\beta}) u \bar{\delta} \sigma; u}) +2 u_{(\bar{\alpha}}(R_{\bar{\beta}) \bar{\gamma} \bar{\delta} u; \sigma}-2(R_{\bar{\beta}) \bar{\gamma} \bar{\delta} \sigma; u}) +3(R_{\bar{\alpha} \bar{\gamma} \bar{\beta} u;u } +R_{\bar{\alpha} u\bar{\beta}  \bar{\gamma} ;u })\sigma_{\bar{\delta}} +3 R_{\bar{\gamma}\bar{\delta} u \sigma ;u } u_{\bar{\alpha}}u_{\bar{\beta}}\right) \right].
\end{align}
Here indices labelled with with a $\sigma$ or $u$ are contracted with $\sigma^{\alpha}$ or $u^{\alpha}$ respectively. For example,  $R_{u \sigma u \sigma}=R_{\bar{\alpha} \bar{\beta} \bar{\gamma} \bar{\delta} } u^{\bar{\alpha}} \sigma^{\bar{\beta}} u^{\bar{\gamma}} \sigma^{\bar{\delta}}$.
\end{widetext}

\section{Useful properties of the Dirac delta distribution}
\label{sec:deltaapp}
As this work involves derivatives of the Dirac delta, the idea of the derivative must be extended to these distributional functions. By integrating against a test function, one can show that the derivative of the Dirac delta satisfies the property
\begin{equation}
\label{deltaprop2}
f(x)\delta'[x-a]=f(a)\delta'[x-a]-f'(a)\delta[x-a].
\end{equation}
Properties of higher derivatives of the delta distribution may be found through the direct differentiation of equation~\eqref{deltaprop2}, such that the second derivative of the delta distribution satisfies
\begin{equation}
\label{deltaprop3}
f(x)\delta''[x-a]=f(a)\delta''[x-a]-2f'(a)\delta'[x-a] +f''(a)\delta[x-a]. 
\end{equation}

The composition of a Dirac delta distribution with with a smooth and continuously differentiable function, $g(x)$, satisfies
\begin{equation}
\label{deltacomposition}
\delta[g(x)]=\sum_{i} \frac{\delta\left[x-x_{i}\right]}{\left|g^{\prime}\left(x_{i}\right)\right|},
\end{equation}
where $x_{i}$ are the roots of $g(x)$ and it is assumed that the roots are simple and $g'(x_i)\neq 0$. Differentiating equation~\eqref{deltacomposition} and making use of \eqref{deltaprop2} yields the equivalent property for the composition of the derivative of the Dirac delta function with a function $g(x)$,
\begin{equation}
\label{deltaprimecomp}
\delta'[g(x)]=\sum_{i} \left(\frac{\delta'\left[x-x_{i}\right]}{\left|g^{\prime}\left(x_{i}\right)\right| g'(x_i)} + \frac{\delta\left[x-x_{i}\right] g''(x_i)}{\left|g^{\prime}\left(x_{i}\right)\right| g'(x_i)^2} \right).
\end{equation}
As before it is assumed that the roots are simple and $g'(x_i)\neq 0$.\\

\section{Monopole metric completion in the Zerilli gauge}
\label{sec:lorenzmono}

In this appendix we summarize the derivation of the $\ell=0$ perturbation presented in Sec.~\ref{sec:metric_reconstruction}. Solving the linearized and mode decomposed Einstein Field Equations directly for $\ell=0$ in a Zerilli-like gauge, we obtain the two non-zero monopole metric perturbation components for an aligned-spin secondary in a circular equatorial orbit in Schwarzschild spacetime:
\begin{align}
\label{eqn:generalmonopole1}
h_{rr}=&\frac{2 \mu ^2 K_{3}^{tt}   \delta [r-r_p]}{rf}+\frac{2 \mu  E\Theta [r-r_p]}{rf^2}+\frac{c_1}{rf^2},\\
h_{tt}=&\frac{2 \mu E \Theta [r-r_p]}{r } \frac{rf}{r_pf_p}\left[\frac{r_p-r}{rf}+\mu \chi  \Omega   \frac{2 r_p -3M}{r_p-2M} \right]\nonumber \\
\label{eqn:generalmonopole2}
&+c_2 f+\frac{c_1}{r},
\end{align}
where $c_1$ and $c_2$ are constants of integration. Note that these expression are valid in either the fixed frequency or fixed radius parameterisation by substituting the corresponding $E$ and $r_p$ of either parameterisation and expanding through linear order in spin.\\

By `Zerilli-like' gauge, we mean a gauge in which the trace of the metric perturbation on the unit two-sphere vanishes; $K\equiv 1/(2r^2)(h^{\theta \theta} + \sin^{2}\theta h^{\phi \phi})=0$. $c_2$ characterizes the residual freedom within this gauge after imposing the additional gauge conditions $h^{\ell=0}_{tr}=\partial_t h^{\ell=0}_{\alpha\beta}=0$. To see this, first note that a monopole gauge vector has the form $\xi^{\alpha}=\{\xi^t, \xi^r,0,0\}$. Under a gauge transformation, $K$ changes by $\delta_{\xi}K=-\frac{2}{r}f \xi_r$ and thus fixing $K=0$ fixes $\xi_r$. Requiring the monopole to be static requires fixing $h_{tr}=0$ which changes as $\delta_{\xi}h_{tr}=-\partial_r\xi_t +\frac{2M}{r^2 f}\xi_t$ under a gauge transformation, limiting $\xi_t$ to be of the form $\xi_t= g(t) f(r)$. Finally as $\delta_{\xi}h_{tt}=-2\partial_t\xi_t$, a static monopole requires that the function $g(t)$ is of the form $g(t)=-\frac{t*C}{2}$ where $C$ is a constant. Then, $\delta_{\xi}h_{tt}=C f(r)$ and with the identification $C=c_{2}$, the remaining gauge freedom is captured by the choice of $c_2$ in \eqref{eqn:generalmonopole2}. 

The constant $c_1$ is not a gauge freedom --- we have set $c_1=0$ to ensure the monopole perturbation has the correct mass (it is a nice addition that this also ensures the monopole is regular at the horizon). By correct mass, we mean that every sphere of radius $r<r_p$ contains a mass $M$; for $c_1\neq0$, the mass enclosed by such spheres is instead $M+c_1/2$, meaning the background mass $M$ differs from the black hole's physical mass $M+c_1/2$. Since the mass is gauge invariant~\cite{Dolan:2012jg}, it is independent of $c_2$.\\

The asymptotic behaviour of the metric perturbation is
\begin{align*}
\lim_{r\to\infty} h_{tt} & = c_ 2 - \frac {2 \mu E}{r_pf_p}  \left [1- \mu \chi  \Omega   \frac{2 r_p -3 M}{r_p-2M}\right],\\
\lim_{r\to\infty} h_{rr} & = 0,
\end{align*} 
and requiring asymptotic flatness restricts our choice to a unique $c_2$ :

\begin{equation}
\label{eqn:flat}
 c_ 2 = \frac {2 \mu E}{r_pf_p}  \left [1- \mu \chi  \frac{2 r_p -3 M}{r_p-2M}  \Omega  \right].
\end{equation} 

Selecting the gauge (i.e., $c_2$) to impose asymptotic flatness and expanding the monopole to linear order in spin with $r_p=r_\Omega + r_\chi$, we obtain the monopole listed in equation~\eqref{eqn:monopole}. In this form we have the retarded monopole in a Zerilli-like gauge that is static, well behaved at the horizon, asymptotically flat and has the correct mass-energy. The linear-in-spin discontinuity is a result of the spin-dipole contribution to the Detweiler Whiting singular field; the regularised monopole is continuous.  We have opted to use this particular monopole in our numerical calculations so that we may check our redshift results with those of Ref.~\cite{Bini:2018zde}, who first derived and used this monopole for a spinning secondary.

\bibliography{SpinningSecondary}

\end{document}